\documentclass[aps,pre,twocolumn,showpacs,amsmath,amssymb,reprint]{revtex4-1}

\usepackage{graphicx}
\usepackage{dcolumn}
\usepackage{bm}

\newcommand{\fo}{\widehat{\mathcal{L}}}
\newcommand{\bo}{\widehat{B}} 
\newcommand{\ko}{\widehat{\mathcal{K}}} 
\newcommand{\lo}{\widehat{L}}

\begin{document}


\title{Hard-Ball Gas as Hard Nut of Statistical Mechanics \\ %
(why mathematicians missed 1/f-noise there)}

\author{Yu. E. Kuzovlev}
\affiliation{Donetsk Institute for Physics and Engineering, Donetsk}
\email{kuzovlev@fti.dn.ua}


\begin{abstract}
We continue discussion of hard-ball models  %
of statistical mechanics,   %
by example of random walk of hard ball   %
immersed into equlibrium ideal gas.  %
Our goal is to highlight decisive role of  specific  %
phase-space subsets, despite their  %
vanishingly smaall Lebesgue measures under  %
the Boltzmann-Grad limit. The ``art of draining'' such subsets in  %
conventional mathematical constructions resulted in loss of so principal  %
property of many-particle systems  %
as 1/f-noise in diffusivities, mobilities  %
and other transport and relaxation rates. We suggest new  %
approaches to formulation and analysis of evolution equations  %
for hierarchy of probability distribution functions  %
of infinite hard-ball systems,  %
thus further overcoming   %
prejudices of Boltzmannian kinetics and   %
mistakes of its modern adepts.
\end{abstract}



\maketitle




\section{\,Introduction}

{\bf 1}.\, A system of (infinitely) many elastic rigid, or hard, balls  %
(hard spheres) is attractive model of classical gases with  %
short-range repulsive interactions. Especially, - as many do believe, - for %
desired rigorous derivation of celebrated Boltzmann's kinetic equation (BE)    %
under the Boltzmann-Grad limit (BGL). %
Most significant results of the corresponding mathematical activity are collected in  %
monograph \cite{cpg} and earlier review article \cite{pg}.  %

Unfortunately, these results are in contradiction %
to theoretical analysis performed in \cite{i1}, %
moreover, to the N.\,Krylov's fundamental criticism \cite{kr} of prejudices   %
acclimatized in statistical mechanics.  %
These works pointed out why BE  %
has no chances to be valid even under BGL.  %

The essence of this contradiction is very simple. %
On one hand, BE declares a priori definite differential %
cross-section of collisions presuming   %
that they obey uniform (probability)    %
distribution over impact parameter  %
values. On the other hand,   %
in reality there are no physical mechanisms   %
to enforce collisions of any given particle   %
to build up some smooth distribution,  %
all the more a priori predictable one \cite{i1,fn1}.   
Hence, there are no physical grounds for thinking in terms  %
of imaginary a priori ``cross-sections'' or ``probabilities''  %
of (various sorts of) collisions, or other   %
beforehand established characteristics of time  rates    %
of  random events \cite{kr}). 

\,\,\,

{\bf 2}.\, Such radical controversy  %
between two ways of thinking about  %
the same things says that one of them stands on wrong concept  %
or postulate. Below, we shall argue once again that it is the  %
Boltzmann molecular chaos paradigm.  %
It seems so doubtless that itself   %
provokes mistakes in attempts  of its    %
mathematical substantiation expounded in \cite{cpg,pg}.  %
More precisely, formal methods exploited in \cite{cpg,pg}   %
(such as artificial filtration of initial conditions and  %
term-by-term consideration of BGL of infinite  %
iteration series)  %
ignore the fact that actual dynamical roles of different many-particle  %
configurations (clusters and events) are not proportional  %
to their native (Lebesgue or Gibbs) probability measures.  %
As the consequence, many important factors were missed there. 

By these reasons, below we suggest new visual illustrations  %
of statistical  significance  of non-typical (``improbable'')  %
many-particle configurations,   %
even in  BGL, hence, existence of strong statistical correlations  %
between their constituting particles.   

\,\,\,

{\bf 3}.\, Some of defects of conventional formalism  %
are implied by ambiguities in its probability-theoretical   %
formulation of the hard-ball collision rules. 
The matter is that conventional formulation treats collisions like instant states,  %
instead of events with non-zero, let small, duration. 
This, in turn, implies neglecting above mentioned ``improbable''  %
configurations and eventually loss of  ``lion's share''   %
of theory's physical meaning.

On account of all that, below we   %
suggest an alternative probability-theoretical   %
representation of hard-ball collisions, by introducing them  %
as limit case of interactions via continuous  potentials.  
The corresponding non-standard treatment of hard-ball limit of  %
the Bogoliubov-Born-Green-Kirkwood-Yvon (BBGKY)  %
hierarchy of equations is fully consistent with  %
more general considerations \cite{i1,fn1}.  %

This approach, as well as some other original tricks  %
and approaches suggested below,  help to realize that  kinematic    %
possibility of collisions is sufficient reason  %
for appearance of statistical correlations between related particles.    %
This insight can prevent at least a part of wrong  %
hypotheses about ``independence'' of random events..  

\,\,\,

{\bf 4}.\, For brevity and simplicity,  %
we shall concentrate mainly on special but  %
principally important problem of   %
(hard-ball) ``molecular Brownian particle'' in thermodynamically   %
equilibrium ideal gas. At that, we  use some designations and  %
definitions from preprints \cite{hs,hs1}  %
where hard-ball systems already were under our attention.


 \newpage  \begin{widetext}

\,\,\, 

\,\,\,  

 \tableofcontents  \newpage  \end{widetext}  


\section{\,Basic properties of rigid elastic collisions and  %
question of their statistical description}

\subsection{Hard-ball collision rules}

Undoubtedly, in Hamiltonian statistical mechanics this rule  %
must establish that total momentum of pair of particles, -  %
let with masses $m$ and $M$\,, - conserves under their  %
collision, \,$P+p=P^* +p^*$\,, while their relative velocity  %
 \,$u=v -V$\, (with \,$v=p/m$\, and \,$V=P/M$\, being  %
individual velocities) changes according to

\begin{equation}
\Omega\cdot u\,=\,-\Omega\cdot u^*\,\,\,, \label{un}
\end{equation} 

\  with \,$\Omega$\,  denoting unit vector  %
(\,$\Omega =\rho/|\rho|$\,,\, $|\Omega|=1$\,) parallel  %
to radius-vector  \,$\rho =r-R$\, connecting centers of the particles  %
at {\it \,perigee\,} of collision. Specificity of hard balls is that %
there  $|\rho|=a=\,$const\, regardless of $|u|$\,, %
so that at the {\it \,perigee\,} always $\rho =a\Omega$\,.  %
At that, tangential component of $u$ conserves as usually, 

\begin{equation}
(1-\Omega\otimes\Omega)\,u\,=\,(1-\Omega\otimes\Omega)\, u^*\,\,\,, \label{ut}
\end{equation} 

\  and $\Omega$ together with $u$ give complete kinematic  %
characterization of the collision in itself.

\subsection{Conventional probability-theoretical representation  %
of hard-ball collision rules} 

In statistical mechanics, in place of kinematic and dynamic  %
characteristics of particles' motion one has  %
their ``statistical ensembles'' and deals with probability distribution functions (DF). 
 
If elastic collision of hard balls is thought of as %
an ``instant event'' consuming neither time nor space,  %
then it seems natural to represent it by boundary condition  %
for a DF, let $F(\cdot)$\,, as follows: 

\begin{equation}
F(\rho =a\Omega,u)\,=\,F(\rho =a\Omega,u^*)\,\,\,, \label{ccr}
\end{equation}

\  where $u$ and $u^*$ are interpreted as  %
relative velocities ``before'' and ``after'' collision  %
(or vice versa) satisfying relations (\ref{un})-(\ref{ut}). 

This is basic prescription of conventional formal construction %
of hard-ball statistical mechanics (SM) \cite{cpg,pg}. %

\subsection{Confusions of conventional  %
probability-theoretical description of hard-ball collisions} 

Common beliefs in physical adequacy  %
of the condition (\ref{ccr}) in fact is beneath criticism  %
and can be easy destroyed.

\,\,\, 

{\bf 1}.\, Indeed, first of all, let relative velocity   %
before a collision is strictly definite,  %
 \,$u=u_0$\, (with \,$\Omega\cdot u_0<0$\,),  %
so that in respect to it  %
some DF looks like delta-function. Without loss of generality,   %
we can require its  %
normalization to unit. Then, involving also outcome of the collision  %
and following condition Eq.\ref{ccr},  %
we have to write 

\[
F(\rho =a\Omega,u)\,=\,\delta(u-u_0)\,+\,  %
 \delta(u-u_0^*)\,\,\, 
\]

\ This expression, however, evidently violates the normalization.  %
Therefore it should be manually changed to    

\[
F(\rho =a\Omega,u)\,=\,\frac 12\,\delta(u-u_0)\,+\,  %
\frac 12\, \delta(u-u_0^*)\,\,\, 
\]

It thus shows that the Eq.\ref{ccr},  %
under properly corrected interpretation,  %
represents particles which are ``by half before'' and ``by half after''  %
collision. 

\,\,\, 

{\bf 2}.\, This observation reminds that any real  %
collision is not an instant state but a process, or ``event'',  %
more or less extended in space and time.  
In other words, it includes relative motion of colliding particles  %
which, therefore, falls out from motion of collision as the whole.  

Consequently, spatial distribution of number density of collision events %
drifts with centre of mass velocity \,$(MV+mv)/(M+m)$\,.  %
This circumstance, in turn, inevitably implies violation of  %
Boltzmann's molecular chaos, in the form of 1/f-noise  %
in diffusivities (mobilities) of gas particles \cite{i1}. 

Taking into account non-vanishing duration of collisions,   %
and considering them in the centre of mass CM) frame, we may rewrite  %
the condition (\ref{ccr}) as 

\begin{equation}
F(\rho =a\Omega - u\,dt,u)\,=\,  %
F(\rho =a\Omega +u^* dt^*,u^*)\,\,\,, \label{accr}
\end{equation}

\ with an infinitesimal \,$dt>0$\, and   %
\,$dt^*>0$\,. Clearly, this condition,  %
in addition to (\ref{ccr}), prescribes also 

\begin{equation}
-\,[\, u\cdot\nabla_\rho\,F(\rho,u)\,]_{|\rho| =a}\,=\,  %
[\, u^*\cdot\nabla_\rho \,F(\rho,u^*)\,]_{|\rho| =a}\,=\,0\,  \,\, \label{dccr}
\end{equation}

\   It just means that  relative motion of particles  %
is inner part of collision's constitution and thus is excluded  %
from particles' drifts (flights)  during collision \cite{hs1}. 

\,\,\, 

{\bf 3}.\, To continue our critical remarks, notice that  %
from physical viewpoint, strictly speaking,  %
there is no necessity to identify \,$u$\,  and \,$u^*$\,  %
in (\ref{ccr}) with ``initial'' and ``final''  %
values of relative velocity (before and after collision).  %
Instead, one has rights to  %
interpret $F(\rho =a\Omega,u)$ in (\ref{ccr})  %
as probability distribution of intermediate values of relative velocity  %
which can be found in the course of collision (``inside collision''). 

Such vision naturally appears, for instance,  %
when considering hard-ball limit of BBGKY equations.  %
Let us illustrate how it does.

\section{Alternative (non-conventional)  %
probability-theoretical formulation of hard-ball collision rules}

\subsection{Derivation of alternative formulation}

Highlighting, as above, and writing out variables of only one pair %
of interacting particles, we may symbolize evolution  %
of a DF \,$F(\cdot)$\, by equation  

\begin{equation}
\begin{array}{c}
\dot{F}(t,\dots\,\mathcal{E},\rho,P,p\,\dots)\,=\,   %
\\ =\,  %
[\,\dots\,-\, u\cdot\nabla_\rho\,+\, %
\Phi^\prime(\rho)\cdot (\nabla_p -\nabla_P)\, \dots ]\,\times\,  
\\ %
\,\times\, F(t,\dots\,\mathcal{E},\rho,P,p\,\dots)\, \,, \label{see}
\end{array}
\end{equation}

\ where \,$\mathcal{E}= \Phi(\rho)+P^2/2M +p^2/2m$\, is energy of the pair, %
  \,$\Phi(\rho)$\, is  interaction potential, to be short-range, repulsive and  %
 spherically symmetric ($\Phi(\rho)=\Phi(|\rho|)$),  %
 and \,$\Phi^\prime(\rho) = \nabla_\rho \Phi(\rho)$\, is interaction force. 
 
 The dots in Eq.\ref{see} replace omitted terms of evolution operator  %
 (in square brackets), the pair's centre of mass position  %
   $(MR+mr)/(M+m)$  and ``extra particles' '' variables. %

 At the same time, we advisedly introduced %
  argument $\mathcal{E}$\,, as if it was independent on others,   %
in order to get possibility to treat the DF $F(\cdot)$\,, - even  %
at arbitrary sharp  $\Phi(\rho)$\,, - as a smooth function of the partial  %
argument \,$\rho$\, in itself. The smoothness means naturally that  %

\begin{equation}
\frac {|u\cdot\nabla_\rho\,F(\cdot)|}{F(\cdot)}\,<\, \infty\,\,\,, \label{rsm} 
\end{equation}

where gradient $\nabla_\rho$ does not act onto $\mathcal{E}$\,. 

The hard-ball interaction results from infinitely sharp %
potential when 

\begin{equation}
\begin{array}{c}
\Phi(\rho)\,\Rightarrow\, \infty\,\,\,\,\, \texttt{if}\,\,\,\,\, %
|\rho|< a\,\,\,, \\ 
\Phi(\rho)\,\Rightarrow\, 0\,\,\,\,\, \texttt{if}\,\,\,\,\, %
|\rho|\geq a\,\,\,
\end{array}    \label{hbl}
\end{equation}

\ We thus take in mind a sequence of system's evolutions corresponding %
to sequence of interaction potentials tending to the hard-ball one.  %
At that, we require that at any of these evolutions  %
any DF $F(\cdot)$ stays a smooth function of the  %
evolution time as well (except, may be, very initial stage of evolution,  %
but certainly  at late enough ``kinetic'' stage,   %
in Bogolyubov's terminology \cite{bog}).  %
This means that, similarly to (\ref{rsm}),     

\begin{equation}
\frac {\,|\partial_t {F}(\cdot)|}{F(\cdot)}\,<\, \infty\,\, \label{tsm} 
\end{equation}

\  Then from Eq.\ref{see}, together with identity  %

\[
[\, -\,u\cdot\nabla_\rho\,+\,  %
\Phi^\prime(\rho)\cdot (\nabla_p -\nabla_P)\,]\,\,  %
\mathcal{E}(\rho,P,p)\,=\,0\,\,\,, \, 
\]

it follows that necessary boundedness  

\begin{equation}
\frac {\Phi^\prime(\rho)\cdot (\nabla_p -\nabla_P)\, %
F(\cdot)}{F(\cdot)}\,\Rightarrow\,  %
\gamma(\cdot)\,\neq\, \infty\, \, \label{req}
\end{equation}

\  also takes place under limit transition (\ref{hbl}),  %
with differentiation operators (momenta gradients),  %
  \,$\nabla_p$\, and \,$\nabla_P$\,,  %
acting on the \,$F(\cdot)$\,'s momentum  %
arguments in themselves only (i.e. not touching \,$\mathcal{E}$\,).  %

Clearly, all this means that under the transition   

\begin{equation}
\frac {\Omega\cdot (\nabla_p -\nabla_P)\, F(\cdot)}  %
{F(\cdot)}\,\Rightarrow\,0\,  %
\,\,\,\,  \texttt{at}\,\,\,\,\, %
|\rho| \leq a\,\,\,,    \label{req_}
\end{equation}

\ where \,$\Omega =\rho/|\rho|$\,, and %
operators \,$\nabla_p$ and $\nabla_P$ do not touch the energy  %
factor\, $\mathcal{E}$\,.  
Or, equivalently, - since region $|\rho|<a$ may be considered as forbidden after %
the hard-ball limit transition, -  

\begin{equation}
\Omega\cdot (\nabla_p -\nabla_P)\, F(\rho =a\Omega)\, %
\,=\,0\,  \,\,,    \label{cr}
\end{equation}

\ again with  $\nabla_p$ and $\nabla_P$\, ignoring factor \,$\mathcal{E}$  %
 (which, of course, now turns to mere  %
kinetic energy). 

\,\,\,

The Eq.\ref{cr} thus must serve in place of Eq.\ref{ccr}  %
in the role of boundary condition, %
in the \,$r-R=\rho$\,\,-space, for probability density evolution equations.  
In other words, Eq.\ref{cr} gives alternative to (\ref{ccr})   %
non-conventional probability-theoretical representation of hard-ball  %
collision rules. 

\subsection{Discussion. The alternative probabilistic formulation  %
of hard-ball collision rules directly forbids  Boltzmann's molecular chaos}

{\bf 1}.\, One could see that just suggested unusual  %
boundary condition (\ref{cr}) is logically implied by  %
very simple and physically meaningful mathematical reasonings,  %
in contrast to the traditional condition (\ref{ccr}) which   %
was merely postulated somewhere as naive %
literal reflection of mechanical relations (\ref{un})-(\ref{ut}). 

The surface appearances of usual and alternative  %
 conditions also are quite different.  %
Nevertheless, there is no qualitative disagreement between  %
their mathematical contents. Indeed, notice, first,  %
that factor $\mathcal{E}$  %
in the DF  \,$F(t,\dots\,\mathcal{E},\rho,P,p\,\dots)$\, in (\ref{cr}), -  %
  \,$\mathcal{E}=P^2/2M +p^2/2m$\, after the hard-ball limit, -  %
is invariant in respect to changing \,$u$\,  %
to \,$u^*=u-2\Omega (\Omega\cdot u)$\,,  %
in accordance with (\ref{un})-(\ref{ut}).
Second, if we represent the same DF through variables \,$P+p$\,  %
and \,$u=v-V=p/m-P/M$\, instead of \,$P$\, and \,$p$\,, then condition (\ref{cr})  %
says that 

\begin{equation}
\Omega\cdot \nabla_u\,  %
F(t,\dots\,\mathcal{E},\rho =a\Omega,P+p,u\,\dots)\,  %
\,=\,0\,  \,    \label{cr_}
\end{equation}

\ at fixed $\mathcal{E}$\,.  Hence, \,$F(\rho =a\Omega)$\, has no   %
dependence on normal component of relative velocity, \,$\Omega\cdot u$\,, at all.  %

As the consequence, obviously, summary DF's dependence on \,$u$\,  %
satisfies\, $\,F(\rho =a\Omega, u) =F(\rho =a\Omega, u^*)$\,,  %
that is our condition (\ref{cr}) contains  %
conventional condition (\ref{ccr}). 

\,\,\,

{\bf 2}.\,\,But the opposite statement generally is wrong.  %
Therefore, our condition is more restrictive and may forbid somewhat   %
allowed by the conventional one.  %

Anyway it is easy to make sure that in general   %
the Boltzmann's ``molecular chaos'',  %
that is pair DF's factorization for particles entering a collision  %
(at $\rho\rightarrow a\Omega$\,, $\Omega\cdot u <0$\,),  %
certainly is forbidden. Indeed, according to the above derivation  %
of our boundary condition (\ref{cr}), most general factorized DF   %
can be expressed by 

\[
F(|\rho|=a)\,=\,e^{-\beta \mathcal{E}}\,  %
A(R,P)\,B(R+a\Omega,p)\,\,\,, 
\]

\ with some coefficient \,$\beta$\,,\, and \,$\mathcal{E}$\, not subject  %
to the differentiation operators  %
in Eqs.\ref{cr}-\ref{cr_} (thus acting onto product $A\,B$ only).  %
Then Eq.\ref{cr} requires 

\[
\Omega\cdot\nabla_P\,\ln\,A(R,P)\,-\,  %
\Omega\cdot\nabla_p\,\ln\,B(R+a\Omega,p)\,=\,0\,\,\, 
\]

\ at any $\Omega$\,. This is possible only when  %
  \,$A\,\propto\,\exp{(c\cdot P)}$\, and \,$B\,\propto\,\exp{(c\cdot p)}$\,  %
with one and the same constant vector \,$c$\,   %
and (omitted) proportionality coefficients  %
depending respectively on \,$R$\, and \,$r$\, only. %

Hence, the factorization is compatible with Eq.\ref{cr} in the only case  %
when momenta distribution is thermally equilibrium Maxwellian one  %
(may be shifted in velocity space by \,$c/\beta$\,), %
i.e. when collisions make no effect at all. 

\,\,\, 

{\bf 3}.\, Thus, interestingly, in the framework   %
of our probability-theoretical  %
representation of hard-ball collision rules, any particles  %
taking part in mutual collision (at least factually resultant one)  %
possess significant mutual statistical correlations.  
Moreover,  these correlations nearly equally cover  %
both post-collision ($\Omega\cdot u>0$) and pre-collision ($\Omega\cdot u<0$)  %
configurations. 

One can say that intervention of particles  %
in same collision is sufficient cause  %
for statistical inter-correlations between them.  %
Characteristic ``mathematical mechanisms''  of creation of these correlations  %
and their physical meaning were pointed out already in \cite{i1} %
and then many times discussed in other works \cite{fn1}. 
 
Our present consideration newly shows how  %
collisions-induced inter-particle correlations can manifest  %
themselves even at level of separate collision event and single evolution equation. 
But, of course, in order to investigate conjugated  %
statistics of actual random series of collision events,  %
we need in full infinite BBGKY hierarchy. 


\section{Hard-ball Brownian particle in ideal gas}

First, let us recollect general case of smooth interaction potential. 

\subsection{BBGKY hierarchy for a smooth interaction potential} 

{\bf 1}.\, The BBGKY hierarchy, which describes  %
``molecular Brownian particle'' (BP)  %
interacting with atoms of ideal gas, can be written as 

\begin{equation}
\dot{F}_k\,=\,-V\cdot\nabla_R\,F_k -\sum_{j=1}^k\,\widehat{L}_j\, F_k\, -  
\,n\nabla_P \int_{k+1}\!\!\Phi^{\,\prime}(\rho_{k+1}) \,F_{k+1}\,  %
\, \label{fn}
\end{equation}

\ Here $\,k=0,1, ...\,$ is number of gas atoms under simultaneous %
attention along with BP;\, $\,F_{k}\,$ is corresponding\, $(k+1)$-particle  %
DF;\, $\widehat{L}_j$ is Liouville operator  %
describing motion of $j\,$-numbered atom  %
(atoms) and its (their) interaction with BP,

\begin{equation}
\widehat{L}_j\,=\, -\,u_j\cdot\nabla_{\rho_j}\, +\,  %
\Phi^\prime (\rho_j)\cdot [\nabla_{p_j} -\nabla_P\,]\,\,\,; \label{lj}
\end{equation}

\ $\rho_j =r_j -R$\,, $u_j =v_j-V$\,, $v_j =p_j/m$\,, $V=P/M$\,;\,  
 $\int_s \dots\,=\,\int \! \int \dots\, d\rho_s\,dp_s$\,,\,   %
 and $n$ is mean gas density. 

\,\,\,
 
We are interested first of all in BP's random walk  %
in thermodynamically equilibrium gas. Therefore,  %
initial conditions to Eqs.{fn} will be 

\begin{equation}
F_k|_{t=\,0}\,=\,\label{ic}   %
\delta(R)\,G_M(P) \prod_{j\,=1}^k \,g(x_j)\, \,\,, \, 
\end{equation}

\  with notations\, \,$x =\{\rho,p\}$\,,  

\[
g(x)=\,E(\rho)\,G_m(p)  \,\,\,, \, 
\]

\  and  
\[
E(\rho)\,=\,\exp{[-\Phi(\rho)/T\,]}\, \,\,, \,\,\,\,\,  
\]
\[ 
G_m(p)\,=\,(2\pi Tm)^{-3/2}\exp{(-p^2/2Tm)}\,\, 
\]
(thus $\,G_m({\bf p})\,$ denoting Maxwell   %
momentum distribution of a particle with mass $\,m\,$).  
 
Clearly, the corresponding  %
 DFs describe BP which at $t<0$ was fixed near the  %
coordinate origin, being surrounded by equilibrium gas, but   %
at $t=0$ becomes released. The release destroys statistical equilibrium  %
(detailed balance) between BP and gas and  %
initiates transition of the system to new equilibrium (new detailed balance)  %
where BP's position will be fully uncertain. %
This process creates  specifically   %
non-equilibrium  many-particle  statistical    %
correlations between BP and atoms. 
Full hierarchy of these correlations serves as ``book-keeping report''  %
accumulating information about {\it \,a posteriori\,} probabilities  %
(actual statistical weights) of various BP's collision patterns  %
and resulting  trajectories. 

To solve Eqs.\ref{fn}, we have also to take into account the  %
trivial boundary conditions for DFs at infinity:\,  
$\,F_k\,\rightarrow\,F_{k-1}\,G_m({\bf p}_s)\,$\,  %
at \,$\,\rho_s \rightarrow\infty\,$\,, where $\,1\leq s\leq k\,$ and
$\,F_{k-1}\,$ does not include $\,\rho_s\,$ and $\,p_s\,$.

\,\,\, 

{\bf 2}.\,\,It may be useful to recollect method of  generating functionals (GF)  %
of DFs, - for the first time introduced by Bogolyubov in \cite{bog}, -  %
and exploit  so-called ``dynamical virial relations'' (DVR)  %
for the first time previewed in \cite{p0710} and then  %
substantiated and investigated in  %
\cite{p0802,p0803,p0806,tmf,ig,p1105,p1203,p1209}. 

Here, let us introduce GF by 

\begin{eqnarray}
\mathcal{F}\{t,R,V,\psi\,;\,n\,\}\,= %
F_0\,+  \label{gf} 
\\ + %
\sum_{s=1}^\infty \frac {1}{s!} %
\int_1 \! \dots \! \int_s F_s\, %
\prod_{j=1}^s\,\psi(x_j)\,   
\nonumber
\end{eqnarray}

\ with formally arbitrary probe function $\psi(x)\,$. %
This GF obeys evolution equation 

\begin{eqnarray}
\dot{\mathcal{F}}\,=\,-V\cdot\nabla_R\,\mathcal{F} \,+  %
\label{gfe}   %
\int_x\, [\,n\,+\,\psi(x)]\, \times\, \\ \times\,     
[\,(V-v)\cdot\nabla_\rho +\Phi^\prime(\rho)\cdot  %
(\nabla_p - \nabla_P)]\,  %
\frac {\delta \mathcal{F}}{\delta \psi(x)}\,  \,, \nonumber 
\end{eqnarray}

\  which is equivalent to the whole hierarchy (\ref{fn}),  %
with initial condition 

\begin{eqnarray}
\mathcal{F} (t=0)\,=\,\delta(R)\,G_M(P) \,\, 
\exp\, \int_x g(x)\,\psi(x) \,\,\,, \, \label{fic}   %
\end{eqnarray} 

\ equivalent to all (\ref{ic}). 

What is for the DVR, for our particular system ``BP in ideal gas''   %
under initial conditions (\ref{ic})  %
we can express them e.g. in the form pointed out in \cite{p1209}, 

\begin{eqnarray}
\frac {\partial F_s }{\partial n}\,= %
\int_{s+1} [\,F_{s+1} - g(x_{s+1})\,F_s\,]\, %
\,\, \label{fdvr} %
\end{eqnarray} 

\  Or, equivalently, in terms of the GF, 

\begin{eqnarray}
\frac {\partial \mathcal{F}}{\partial n}\,= %
\int_x \left[\,\frac {\delta}{\delta \psi(x)}\,-\,g(x)\,\right]\, %
\mathcal{F}\,\, \label{gdvr} %
\end{eqnarray}

 Notice that these DVR are valid also for arbitrary non-equilibrium  %
initial gas states represented by any reasonable choice  %
of the function \,$g(x)$\,  different from the above concretized one. 


\subsection{Conventional BBGKY hierarchy for hard-ball interaction} 

Following postulates of the conventional mathematical theory \cite{cpg,pg},  %
in case of hard-ball BP-atom interaction the BBGKY Eqs.\ref{fn}  %
should be replaced by

\begin{eqnarray}
\dot{F}_k\,=\,-V\cdot\nabla_R\, F_k + %
\sum_{j=1}^k\,(V-v_j)\cdot\nabla_{\rho_j}\, F_k\, +  %
\label{hfn} \\ +\,   
\,na^2 \oint \!\int dp_{k+1}\,\, (\Omega\cdot(v_{k+1}-V)) \,  %
\times\, \nonumber \\ \times \,
\, F_{k+1}(\rho_{k+1}=a\Omega)\,\,\,, \nonumber
\end{eqnarray}

\  where \, $\,\oint \dots\,=\,\int \dots\,d\Omega$\,,  %
and $\,|\rho_j|> a$\,. These equations must be supplied  %
by boundary conditions like (\ref{ccr}), 

\begin{equation}
F_k(\rho_j=a\Omega_j,V,v_j)\,=\,  %
F_k(\rho_j=a\Omega_j,V^*, v_j^*)\,  \,\,    \label{ccrs}
\end{equation}

\  or, in terms of variables \,$v_j-V\equiv u_j$\,  %
and \,$P+p_j$\,,  in view of the conservation \,$P+p_j=P^*+p_j^*$\,,, 

\begin{equation}
F_k(\rho_j=a\Omega_j, u_j)\,=\,  %
F_k(\rho_j=a\Omega_j, u_j^*)\,  \,\,,    \label{ccrs_}
\end{equation}

\ where $\,j=1\dots k\,$ and $\,|\Omega_j|=1$\,. 

At that, initial conditions corresponding to (\ref{ic})  %
are 

\begin{equation}
F_k|_{t=\,0}\,=\, \delta(R)\,G_M(P)  %
\prod_{j=1}^k\,G_m(p_j)\,\,\,, \label{hic}
\end{equation}

\  while the conditions (of weakening of correlations)  %
at infinity are  %
  $\,F_k\,\rightarrow\,G_m(p_j)\,F_{k-1}\,$  %
  at \,$\,\rho_j \rightarrow\infty\,$\,,  %
  with $F_{k-1}$ independent on $\rho_j\,$ and $p_j\,$.


\subsection{Alternative BBGKY hierarchy for the hard-ball limit} 

Our above considerationI prompts that before performing the  %
hard-ball limit (\ref{hbl}) in Eqs.\ref{fn} it is necessary  %
to extract from the DFs\, the ubiquitous thermodynamical factors and   %
write

\begin{eqnarray}
F_k\,=\, \exp{(-\mathcal{E}_k/T)}\, Q_k\,=\,  \label{fq}  %
\,\,\,\, \\ \,=\, \{\,G_M(P) \prod_{j=1}^k\, E(\rho_j)\,G_m(p_j)\,\}\,   %
\times\,Q_k\,\,\,,  \nonumber
\end{eqnarray}

\ where \,$\mathcal{E}_k$\, is energy of ``BP plus \,$k$\, atoms''.  %
Such defined functions \,$Q_k$\,  just represent  %
the mentioned perturbations of detailed balance  %
and related non-equilibrium statistical correlations. 

Then, we must take into account that, naturally, characteristic energies  %
(per one particle), conjugated with these perturbations and correlations,  %
remain finite under the limit (\ref{hbl}). Therefore,  %
all the functions $Q_k$ (with $k>0$) remain continuous smooth functions   %
of the distances $\rho_j$\,, in the sense that formally all these functions  %
stay continuously extendable into regions $|\rho_j|<a$\,.

In such way we come to the hard-ball limit scheme formulated %
in Section III. Applying its collision bounary condition, Eq.\ref{cr},\,  %
in Eqs.\ref{fn}, together with Eqs.\ref{fq},   %
after elementary manipulations and reasonings  %
it is not hard to arrive to equations

\begin{eqnarray}
\dot{Q}_k\,=\,-V\cdot\nabla_R\, Q_k + %
\sum_{j=1}^k\,(V-v_j)\cdot\nabla_{\rho_j}\, Q_k\, +  %
\label{qn} \\ +\,   
\,na^2 \oint \!\int dp_{k+1}\,\, (\Omega\cdot(v_{k+1}-V)) \,  %
\times\, \nonumber \\ \times \,
G_m(p_{k+1})\,\, Q_{k+1}(\rho_{k+1}=a\Omega)\,\,\,, \nonumber
\end{eqnarray}

\  again with \, $\,\oint \dots\,=\,\int \dots\,d\Omega$\,,\,   %
and $\,|\rho_j|\geq a$\,,\, %
but now to be supplied by  boundary conditions very visually  %
different from (\ref{ccrs}). Namely,  

\begin{equation}
\Omega_j\cdot (\nabla_{p_j} -\nabla_P)\,  %
Q_k(\rho_j =a\Omega_j)\, \,=\,0\,  \,\,,    \label{crs}
\end{equation}

\ where $\,j=1\dots k\,$ and $\,|\Omega_j|=1$\,. 

\,\,\, 

This is our hard-ball limit case of the BBGKY hierarchy %
for ``BP in ideal gas''. Clearly, the mentioned initial conditions to it %
now look as 

\begin{equation}
Q_k|_{t=\,0}\,=\, \delta(R)\,\,\,, \label{ic_}
\end{equation}

\  while the conditions (of weakening of correlations)  %
at infinity now state that  %
  $\,Q_k\,\rightarrow\,Q_{k-1}\,$ at \,$\,\rho_j \rightarrow\infty\,$\,,  %
  with $Q_{k-1}$ free of $\rho_j\,$ and $p_j\,$.


\subsection{Hard-ball BBGKY hierarchy in terms of  %
cumulant (correlation) distribution functions} 

{\bf 1}.\, Just presented equations (\ref{qn})   %
by themselves  %
have no essential difference from equations (\ref{hfn}) of %
conventional theory. Indeed, the latter turn to the former after  %
replacement 

\begin{equation}
F_k\,=\, \{\,G_M(P) \prod_{j=1}^k\, G_m(p_j)\,\}\,   %
\times\,Q_k\,\,\,,  \label{hfq}
\end{equation}

\ so that the only actual difference between alternative and conventional   %
formalisms is in their collision (contact) boundary conditions.  %
Namely, our ones are given by Eqs.\ref{crs}, while conventional,    %
Eqs.\ref{ccrs}, - when considered in terms of  \,$Q_k$\,  from (\ref{hfq}), -  %
appear from  Eqs.\ref{ccrs} by mere inserting  %
  \,$Q_k$\, in place of \,$F_k$\,. 

Hence, the concept of {\it \,cumulant\,}, or {\it \,correlation\,},   %
functions (CF) \cite{hs,fn2} directly transmits to our formalism.  %
Designating them by \,$C_k$\,, as in \cite{hs},   %
and introducing like there, but now through \,$Q_k$\,'s instead of \,$F_k$\,'s,  %
we have to write 

\begin{eqnarray}
Q_0(t,R,P)= C_0(t,R,P)\,\,\,, \,\,\,\,\,\,\, \nonumber \\ 
Q_1(t,R,P,x_1)= C_0(t,R,P)+ C_1(t,R,P,x_1) \,\,\,, \,\,\,\, \label{hqc}\\ 
Q_2(t,R,P,x_1,x_2)\,=\,C_0(t,R,P)\,+ \, \nonumber\\ 
+\, C_1(t,R,P,x_1)\,+ \,C_1(t,R,P,x_2) \,+\, \nonumber \\  %
+\,C_2(t,R,P,x_1,x_2) \,\,\,, \nonumber
\end{eqnarray}

\ and so on. 

Clearly, advantage of such defined ``cumulant functions''  %
(CF) \,$C_k$\, is that they vanish at infinity:\,  %
 \,$ C_k\rightarrow 0$\, at \,$\rho_j\rightarrow\infty$\,.  %
Therefore one can integrate them over relative distances. 
This means that \,$C_k$\, represent most connected, or irreducible,  %
\,$(k+1)$\,-particle correlations.
Correspondingly, initial conditions (\ref{ic_}) in their   %
terms look maximally simple: %

\begin{equation}
C_k|_{t=\,0}\,=\, \delta_{k\,0}\,\delta(R)\,\,\, \label{cic}
\end{equation}

What is for evolution equations fore the CFs,  %
in case of hard-ball interaction they look exactly as Eqs.\ref{qn} %
minus symbol $Q$\,'s replacement by $C$\,:

\begin{eqnarray}
\dot{C}_k\,=\,-V\cdot\nabla_R\, C_k + %
\sum_{j=1}^k\,(V-v_j)\cdot\nabla_{\rho_j}\, C_k\, +  %
\label{cn} \\ +\,   
\,na^2 \oint \!\int dp_{k+1}\,\, (\Omega\cdot(v_{k+1}-V)) \,  %
\times\, \nonumber \\ \times \,
G_m(p_{k+1})\,\, C_{k+1}(\rho_{k+1}=a\Omega)\,\,\,\, \nonumber
\end{eqnarray}

\ However, the collision boundary conditions for these equations  %
essentially differ from (\ref{crs}) or (\ref{ccrs})   %
since now connect CFs with two neighboring numbers.  %
Namely, in conventional formalism \cite{hs} 

\begin{eqnarray}
C_k(\rho_j =a\Omega_j,P,p_j) + C_{k-1}(P) \,=\,  %
\nonumber \\ \,=\, %
 C_k(\rho_j =a\Omega_j,P^*,p_j^*) + C_{k-1}(P^*) \,\, 
\label{crc}
\end{eqnarray}

\ while in alternative formalism

\begin{equation}
\Omega_j\cdot (\nabla_{p_j} -\nabla_P)\,  %
[\,C_k(\rho_j =a\Omega_j,P,p_j) + C_{k-1}(P) \,]\,=\,0\,\,\,  \label{rc}
\end{equation}

\  These formulas follow from the above CFs definition (\ref{hqc})  %
as applied to (\ref{ccrs}) or (\ref{crs}).  %
In both them  \,$C_{k-1}$\, does not concern \,$j$\,-th atom,  %
and we omitted all arguments not concerned by a collision under consideration.

\,\,\, 

{\bf 2}.\, In the hard-ball limit the DVR (\ref{fdvr}) yield 

\begin{eqnarray}
\frac {\partial Q_s }{\partial n}\,= %
\int_{\rho\,:\, |\rho|>a} \int_{p} G_m(p)\,  %
[\,Q_{s+1} - Q_s\,]\,\,\,,   \nonumber  
\\  %
\frac {\partial C_s }{\partial n}\,= %
\int_{\rho\,:\,|\rho|>a} \int_{p} G_m(p)\,  %
C_{s+1}\,\,\,, \,\,\, \label{qcdvr} %
\end{eqnarray} 

\    with \,$\rho = \rho_{s+1}$\, and \,$p=p_{s+1}$\,.   %

Importantly, these relations hold  %
regardless of choice of collision boundary conditions.  %
The latter  circumstance is due to fact that generally DVR  %
are insensible to character of interactions. %
This is because DVR are expression of general kinematic properties  %
of (infinitely) many-particle dynamical systems,  %
first of all, the phase volume conservation there (expressed also  %
by the ``Liouville theorem'' and ``fluctuation-dissipation   %
relations'' \cite{ufn1}). 

\,\,\, 

{\bf 3}.\, One can see that  irreducible correlations of given order  arise  %
either from lower-order correlations, via the collision boundary  %
conditions, or from higher-order correlations, via the  %
``collision integrals''. Of course, at initial stage of evolution %
the first of these two opposite flows of correlations is dominating.  %
But later, at kinetic stage, their approximate balance  %
may be expected. It then establishes some  spatial bounds    %
correlated clusters, so that \,$C_k$\,'s extension in  %
\,$\rho_j$\,-spaces  is not growing unboundedly   %
with time. 

According to theorem, or ``lemma'', proved in  %
\cite{p0802,p0803,tmf}, such behavior of inter-particle  %
correlations means presence of time-scaleless 1/f\,-type  %
fluctuations of BP's diffusivity (mobility \cite{bk3,bk12}). 
A simple substantiation of this statement, basing on simplest  %
of the DVR, was demonstrated in \cite{ufn1,p1209}.  


\subsection{Hard-ball limit of generating functional equations   %
and dynamical virial relations}

In terms of generating functional (GF) introduced by 

\begin{eqnarray}
\mathcal{Q}\{t,R,V,\psi\,;\,n\,\}\,=\, %
Q_0\,+  \label{gq} 
\\ + %
\sum_{s=1}^\infty\, \frac {1}{s!} %
\int_1 \! \dots \! \int_s Q_s\,\, %
\prod_{j=1}^s\,\psi(x_j) \, \,\,,  
\nonumber
\end{eqnarray}

\ our equations (\ref{qn}) and conditions (\ref{crs}) can be unified into 

\begin{eqnarray}
\dot{\mathcal{Q}}=-V\cdot\nabla_R\,\mathcal{Q} \,+    %
\, \nonumber \\ \,+  %
\int_x \,\,[\,n\,G_m(p) \,\theta(|\rho|-a)  +\psi(x)]\,  %
\times \,\nonumber \\ \,\times\, 
((V-v)\cdot\nabla_\rho) \,  %
\frac {\delta \mathcal{Q}}{\delta \psi(x)}\,\,\,, \,\,\,  \label{gqe}  %
\end{eqnarray}

\  with \,$\theta(\cdot)$\, being the Heaviside step function,  and 

\begin{eqnarray}
\oint \!\int_p\, \psi(a\Omega,p)\, (\Omega\cdot (\nabla_p - \nabla_P))\,  %
\frac {\delta \mathcal{Q}}{\delta \psi(a\Omega,p)}\,=\,0\,\,\, \,\,\label{gcr} 
\end{eqnarray}

\  At that, generating DVR (\ref{gdvr}) transforms to 

\begin{eqnarray}
\frac {\partial \mathcal{Q}}{\partial n}\,= %
\int_{|\rho|>a} \int_{p} G_m(p)\, %
\left[\,\frac {\delta}{\delta \psi(x)}\,-\,1\,\right]\, %
\mathcal{Q}\, \,\,\,\,\,\,\label{gqdvr} %
\end{eqnarray} 

\  It is easy deducable directly from Eq.\ref{gqe}.  %

To rewrite these generating relation in the CF's language,  %
one has to notice that 

\begin{eqnarray}
\mathcal{Q}\{t,R,V,\psi;n\}=\, %
e^{\,\int \psi(x)\, dx}\,  %
\mathcal{C}\{t,R,V,\psi;n\}\,\, ,\,\,  \label{gc} 
\end{eqnarray}

\ where \,$\mathcal{C}$\, is CF's GF introduced similarly to (\ref{gq}).  %


\section{Comparison between standard and alternative  %
treatments of (hard-ball) statistical mechanics}

Although evolution equations in the two approaches are coinciding,  %
their unambiguous solution is impossible without  %
definite collision boundary conditions. But  %
right there the coincidence ends.

\subsection{Why conventional  %
collision boundary conditions seem unsatisfactory}

For the first look, our  %
boundary conditions  %
for probability densities at inter-particle contact  %
surfaces, - (\ref{cr}), (\ref{crs}), (\ref{rc}) and  (\ref{gcr}), -  
are rather complicated and non-transparent   %
in comparison with  %
standard conditions, -  (\ref{ccr}), (\ref{ccrs}) and (\ref{crc}).  
Therefore, it is important to emphasize once more their advantages.  %

In both the Eqs.\ref{hfn} and Eqs.\ref{qn}  %
the ``extra particle'' integral terms, -  %
which eventually must play roles of ``collision integrals'', -   %
are functionals of edge boundary values of DFs at $\,|\rho_{k+1}|=a$\,.   %
Hence, we have all rights, - moreover, are forced, - to be interested  in  %
such edge values of DFs anywhere else besides the integral terms. %
This then requires to consider many-particle configurations where simultaneously  %
  $\,|\rho_{k+1}|\rightarrow a$\, and \,$|\rho_{j}|\rightarrow a$\,,  %
 and so on. Consequently, in general, we need in some boundary conditions   %
 for situations when simultaneously two or several atoms are in contact with  %
 BP or in its arbitrary close vicinity.
 
 That is non-trivial question. Unfortunately, in the context of conventional  %
 theory there is no ready answer to it or a recipe for getting such answer.  
 In any case, one can verify that literal parallel application of two or several  %
 samples of the condition (\ref{ccrs}) can not be a suitable rule  %
 for configurations with two or several \,$|\rho_{j}|\rightarrow a$\, at once, %
 since it is incompatible with conservation of both total momentum and total  %
 kinetic energy of involved particles. 
 
 This fact once again demonstrates that the conventional theory  %
 is formally incomplete. Therefore, there we are enforced to treat  %
 the mentioned configurations as three- or many-particle processes  %
 constituted by two or more almost simultaneous pair collisions.  
 Then one meets extremely complicated task of  %
 geometric and kinematic classification  %
 of infinite variety of such processes. By such reasons,  %
strictly speaking, the conventional  %
 theory seems still rather bad developed.

 \subsection{On problems of probabilistic description of  %
infinitely-many-particle systems}
 
 {\bf 1}.\, In this theory (see e.g. \cite{cpg,pg} and reference therein)  %
 just underlined problems traditionally were avoided, -   %
 taking in mind the Boltzmann-Grad limit (BGL), -  %
 by means of artificial exclusion of ``unpleasant'' configurations  %
 leading to the mentioned many-particle events,  %
 i.e. to sticking together, or ``glued'', pair collisions.  
This is achieved by means of a proper selection of initial conditions  %
under term-by-term consideration of BGL of formal iteration series  %
for BBGKY hierarchy. 
Such the ``art'' is motivated by small statistical weight  %
 (zero in the BGL) of the unpleasant configurations.  
 
However, this is bad idea, because any  %
equation of the BBGKY hierarchy and any DF there   %
represents  \,$1,\,2,\,3\,\dots$\,  particles moving  %
among infinitely   %
many other particles constituting full system,  %
while any term of the iteration series  %
says something about motion of namely  \,$1,\,2,\,3\,\dots$\, %
particles without others. 
At that, actual ``probabilities''  %
of  \,$2,\,3,\,4\,\dots$\,-particle configurations  %
and events in real infinite system  %
hardly are proportional to statistical weights defined for   %
similar configurations and events in a group  %
of fixed finite number of particles. 

{\bf 2}.\, Thus, one should remember that actual statistical effects  %
of collisions are determined by not  {\it \,a priori\,}  %
statistical weights or expectations   %
but {\it \,a posteriori\,} conditional probability densities  %
which reflect both current surroundings of colliding particles and   %
pre-history of the system's evolution. 
For instance, in Eqs.\ref{qn}, by means of factors 
 
\begin{eqnarray}
Q^{\prime}_{k+1}(\Omega,P,p_{k+1}\,|\,t,R,\,x_1\,\dots\,x_k)\,=\,  %
\,\, \nonumber \\ \,=\,
Q_{k+1}(\rho_{k+1}=a\Omega)/Q_k\,\,\, ,\, \,\,\label{sqn} 
\end{eqnarray}

\ which, at \,$\Omega\cdot u_{k+1} <0$\,,  %
visually modify  ``probability of collision'' or  %
``differential cross-section of collision'', - between   %
BP and ``outer'' \,$(k+1)$\,-th atom, -  %
in comparison with what would take place if we considered two  %
isolated particles only in a pre-collision state.  %

In reality, the colliding particles acquire some   %
{\it \,conditional\,} correlation, - in the sense of the probability theory, -  %
conditioned by an information about    %
\,$k$\, other gas atoms  from BP's surroundings. 
The factor (\ref{sqn}),  representing such correlations,  %
more or less differs from unit, \,$Q^{\prime}_{k+1}(\cdot)\neq 1$\,,   %
in particular, if presented information indicates possibility  %
of BP's interaction with some of that \,$k$\, atoms in the past.  
For instance, when  %
the ``outer'' \,$(k+1)$\,-th atom in fact could not arrive closely to BP  %
(to  position with \,$|\rho_{k+1}|\rightarrow a$\,)   %
directly ``from infinity'' (its start position at \,$t=0$) since %
continuation of its straight-line trajectory into the past  %
intersects preceding BP's trajectory as bent by past BP's collision   %
with some of other given  \,$k$\, atoms.   

In the latter case  we meet situation of  ``forbidden'' (or ``impossible''  or   %
``virtual'')  repeated collision, for which we may then suppose that   %
 \,$Q^{\prime}_{k+1}(\cdot)< 1$\,. 
Another particular variant of  %
``unpleasant'' configuration is when it indicates seemingly   %
allowed ``repeated collisions''. Both these examples  %
can be illustrated by figure in \cite{hs}. 

 \,\,\, 

{\bf 3}.\, More complicated cases combine hints of  %
both forbidden and allowed  ``repeated collisions''.  
Of course, relative probability of such events  %
vanishes under the BGL  %
(when \,$a^3n\rightarrow 0$\, while \,$a^2n =$\,const\,)  %
if they are considered from viewpoint of arbitrary initial conditions. 
However, if being considered  from viewpoint  %
of already happened configurations  %
with  \,$|\rho_j|\sim a$\,,  %
they acquire conditional probabilities non-vanishing even in BGL.  
More precisely, if all arguments \,$\rho_j$\, ($j=1\dots k$) in  %
 \,$Q^{\prime}_{k+1}(\Omega,P,p_{k+1}\,|\,t,R,\,x_1\,\dots\,x_k)\,$  %
are kept comparable with \,$a\,$,   %
then, under proper \,$p_j$\, ($j=1\dots k$) ,  %
a portion of  \,$p_{k+1}$\,'s values (weighted  %
with \,$G_m(p_{k+1})$\,), responding to  %
earlier happened interaction between the ``outer''  %
atom and  BP,  stays  comparable with unit  %
(tend to non-zero constant),  %
so that   \,$Q^{\prime}_{k+1}\neq 1$\,, in spite of BGL.  
Hence, ``unpleasant'' configurations may play important,   %
if not decisive, role in true solution to Eqs.\ref{qn}   %
(see \cite{p1203,p1209} and references therein). 

\,\,\, 

{\bf 4}.\,   Another very important thing is   %
that a part of the mentioned   %
``earlier happened'' BP-atom interactions  %
is delegated by the {\it \,seed\,} (equilibrium)  %
BP-atoms correlations represented by function  %
\,$g(x)=G_m(p)\,\exp{[-\Phi(\rho) /T] }\,$,  -  %
or  \,$g(x)\rightarrow G_m(p)\,\theta((|rho| -a)\,$   %
for the hard-ball limit, -  %
in the initial conditions Eq.\ref{ic}.   
 Therefore, formal expression seemingly describing  %
a repeated collision in essence may be description of    %
two stages of one and the same BP-atom collision  %
but statistically influenced by both the seed initial correlations  %
and later generated non-equilibrium correlations  %
due to BP's interaction with the rest of gas. 

The diagram on figure in \cite{hs} applies  %
also to this simple case if we interpret inscriptions  %
\,$C_1^{out}\,$ and  \,$C_1^{in}\,$  there as symbols of  %
``output from initial equilibrium correlation'' and  %
``input to non-equilibrium correlation'', respectively,  %
while \,$C_2$\,  as symbol of  ``influence by the rest of gas'' %
(which just causes a difference  \,$Q^{\prime}_1(\cdot) \neq 1$\,). 

The latter then is not simultaneous intervention of a third  %
particle (atom) but instead interference of previous   %
BP's collisions which altogether  transform ``probability'' or  %
``(differential) cross-section'' of the current collision   %
into random quantity without a priori known average value.  %
More precisely,  with a value whose true  prediction needs in   %
honest solution of the BBGKY hierarchy (e.g.  in terms  of  the  %
factors   \,$Q^{\prime}_{k+1}(\cdot)$\,),. 

Such kind of interference of ``the rest of gas''   %
in particular collision event surely survives under BGL  %
along with corresponding statistical correlations   %
catched in the CFs and \,$Q^\prime $\,s. . 

Unfortunately, these rather fine aspects insensibly  disappear   %
in the framework of conventional theory, because it  operates with  %
initial correlations as if \,$g(x)\rightarrow G_m(p)$\,.  
As the consequence, it incorrectly reproduces action of  %
operators \,$\lo_j$\, onto initial DFs   \,$ F_k(t=0)$\,. 

However, one can avoid such defects if rearranging BGL  %
and hard-ball limit.Other way to correct the theory may be to use %
so-called ``pseudo-Liouville representation''   %
of hard-ball interaction  \cite{cpg,pg} which allows us  %
to unify both the hard-ball    %
BBGKY equations and collision boundary conditions (CBC)   %
(\ref{ccrs}) into single generating functional (GF)   %
evolution equation similar to Eq.{gfe}. 

Therefore it is useful  %
to discuss it, first of all, for smooth interactions,   %
in order to demonstrate character of statistical connections between   %
next BP-atom collision and history of earlier   %
BP's interaction with the rest of gas. 

\subsection{Exact pseudo-kinetic generating-functional   %
formulation of BBGKY hierarchy   %
and crash of Boltzmann's kinetics}

{\bf 1}.\,  Let us introduce functional (differential) operator  

\begin{eqnarray}
\fo\,=\,\fo\{V,\psi,\nabla_P,\delta/\delta\psi\}\,=\, \label{fo}   %
\nonumber \\ \,=\,  %
\int_x\, [\,n\,+\,\psi(x)]\,    %
\lo_x(V,\nabla_P)\,  %
\frac {\delta }{\delta \psi(x)}\,  \,\,, \,\, 
\end{eqnarray}

\  where   \,$\lo_x\,$ is abstraction of operators \,$\lo_j\,$,  

\begin{eqnarray}
\lo_x=\lo_x(V,\nabla_P)\, =\,  \label{lo} \\ \,=\,  \nonumber  %
(V-v)\cdot\nabla_\rho +\Phi^\prime(\rho)\cdot  %
(\nabla_p - \nabla_P)\,\,   %
\end{eqnarray}

\  Besides, it is comfortable to introduce operators 

\begin{eqnarray} 
\fo^\prime\,=\, e^{\,-\int g(x)\, \psi(x)\, dx}\,\fo\,   %
 e^{\,\int g(x)\, \psi(x)\, dx}\,=\, \, \,\,\,\,\,\,\, \label{fop} \\ \,=\,   
\int_x\, [\,n\,+\,\psi(x)]\,   \lo_x(V,\nabla_P)\,  %
\left[\, \frac {\delta }{\delta \psi(x)} \,+   %
\,g(x)\,\right]\,  \,\,, \,\,  \nonumber \\ 
\fo_R\,=\, -V\cdot\nabla_R\,+\,\fo\,\,\,, \,\,\,\,\,\,\,  %
\fo_R^\prime\,=\, -V\cdot\nabla_R\,+\,\fo^\prime\,\,   \nonumber
\end{eqnarray}

\   In parallel, recall definition of the correlation, or cumulant, functions   %
(CF) for general (not hard-ball) BP-atom interaction potential, 

\begin{eqnarray}
F_0(t,R,P)= C_0(t,R,P)\,\,\,, \,\,\,\,\,\,\, \nonumber \\ 
F_1(t,R,P,x_1)= C_0(t,R,P)\,g(x_1) + C_1(t,R,P,x_1) \,\,\,, \,\,\,\, \nonumber \\ 
F_2(t,R,P,x_1,x_2)\,=\,C_0(t,R,P) \,g(x_1)\,g(x_2) \,+ \, \nonumber  \\ 
+\, C_1(t,R,P,x_1) \,g(x_2)\,+ \,C_1(t,R,P,x_2)  \,g(x_1) \,+\, \nonumber \\  %
+\,C_2(t,R,P,x_1,x_2) \,\,\,, \,\,\,\,\, \,\,\,\,\, \label{fc}
\end{eqnarray}

\ and so on, and their generating functional (GF),  
 
\begin{eqnarray}
\mathcal{F}\{t,R,V,\psi;n\}= %
e^{\,\int_x  g(x)\, \psi(x)}\,  %
\mathcal{C}\{t,R,V,\psi;n\}\,\,\,\,  \label{cgf} 
\end{eqnarray}

\  Then the functional evolution equation Eq.\ref{gfe} reads shortly as 

\begin{eqnarray}
\partial_t \mathcal{F}\,=\,-V\cdot\nabla_R\,\mathcal{F} \,+  %
\fo\,  \mathcal{F}\,=\, \fo_R\, \mathcal{F}\,\,\,, \,   \label{gfe_}   %
\end{eqnarray}

\ while equivalent equation for CF's GF as 

\begin{eqnarray}
\partial_t\, \mathcal{C}\,=\,-V\cdot\nabla_R\,\mathcal{C} \,+  %
\fo^\prime \,  \mathcal{C}\,=\, \fo_R^\prime \,  %
\mathcal{C}\,\,\, \,   \label{cfe_}   %
\end{eqnarray}  

\  with initial condition 

\begin{eqnarray}
\mathcal{C} (t=0)\,=\,\delta(R)\,G_M(P) \,\,\, \, \label{fcic}   %
\end{eqnarray} 

\  equivalent to all (\ref{ic})  and  %
independent on the probe-function argument \,$\psi(x)\,$. 

\,\,\, 

{\bf 2}.\,   Next, let us rewrite Eq.\ref{gfe_} in the form
 
\begin{eqnarray}
\partial_t \mathcal{F} = -V\cdot\nabla_R\,\mathcal{F} +      %
\int_x\, [\,n\,+\,\psi(x)]\,    %
\lo_x(V,\nabla_P)\,  %
\mathcal{F}_x  \,  \,\,\, \,\,  \label{gfe1}
\end{eqnarray}

\ where we introduced derivative 
\[ 
\mathcal{F}_x  \,=\, \frac {\delta \mathcal{F}}{\delta \psi(x)} \,\,\,, \, 
\]

and supplement Eq.\ref{gfe1} with equation for  %
\,$\mathcal{F}_x \,$ directly following from Eq.\ref{gfe_}, 

\begin{eqnarray}
\partial_t\, \mathcal{F} _x\,=\,[\,-V\cdot\nabla_R\, +\,\lo_x\,+\,   %
\fo\,\,]\,  \mathcal{F}_x \,\,\,, \,   \label{gfe2}   %
\end{eqnarray}

\  with obvious initial condition 
\[
\mathcal{F}_x(t=0) \, =\,g(x)\, \mathcal{F}(t=0)\,  \, 
\]

Combining all these formulas, it is easy   %
to transform Eqs.\ref{gfe_}  and \ref{cfe_}  to  

\begin{eqnarray}
\partial_t\, \mathcal{F}\,=\,-V\cdot\nabla_R\,\mathcal{F} \,+      %
\,\ko (t)\, \mathcal{F} \,\,\,, \,  \nonumber \\  
\partial_t\, \mathcal{C}\,=\,-V\cdot\nabla_R\,\mathcal{C} \,+   \label{kc}   %
\,\ko^\prime  (t)\, \mathcal{C} \,\,\,, \, 
\end{eqnarray} 

\ with new operators 

\begin{eqnarray}  
\ko(t)\,= \int_x\, [\,n+\psi(x)]\, \lo_x \,  %
\nonumber \,\times \,\,\,\,\,\, \\ \,\times\,\,    %
\exp{[(\lo_x + \fo_R)\,t\,]}\, g(x)\, \exp{[\,- \fo_R\,t\,]}\, \,\,, \,\,   %
\nonumber  \\     
\ko^\prime (t)\,=  \int_x\, [\,n+\psi(x)]\, \lo_x \,  \,\times\,  %
\,\,\,\,\,\,\,\,\, \label{kop}   \\  \,\times \, \, 
\exp{[(\lo_x + \fo_R^\prime )\,t\,]}\, g(x)\, \exp{[\,-  %
 \fo_R^\prime \,t\,]}\, \, \, \nonumber 
\end{eqnarray} 

\  The latter un turn can be transformed like 

\begin{eqnarray}  
\ko^\prime (t)\,= \int_x\, [\,n+\psi(x)]\, \lo_x \,   %
\{\,g(x)+  \int_0^t d\tau\, \partial_\tau \,  
\, \times \,\,, \nonumber   \\  \, \times\,\,  %
\exp{[(\lo_x + \fo_R^\prime )\,\tau\,]}\, g(x)\, \exp{[\,-  %
 \fo_R^\prime \,\tau\,]}\, \}\,  =\,\,\, \nonumber \\ \,=\,    \, \,   
[ \int_x \psi(x)\, \lo_x\,   g(x)\,]\,+\,\,\,\,\,\,\,  %
\,\,\,\, \label{kop1}  \\ \,+ 
\int_x [\,n+\psi(x)]\, \lo_x   \int_0^t d\tau\,  %
\,\times\,\,\, \nonumber \\ \,\times \,\,  %
\exp{[(\lo_x + \fo_R^\prime )\,\tau\,]}\, \lo_x\,  g(x)\,  %
\exp{[\,-  \fo_R^\prime \,\tau\,]}\, \, \, \, \,   \nonumber 
\end{eqnarray} 

\ or, equivalently, 

\begin{eqnarray}  
\ko^\prime (t)\,= \int_x \psi(x)\, \lo_x \,   %
\,\times \,\,\,\,\,\,\, \, \,  \nonumber  \\ \,\times\,\, 
\exp{[(\lo_x + \fo_R^\prime )\,t\,]}\, g(x)\, \exp{[\,-  %
 \fo_R^\prime \,t\,]}\, + \,\,\,\,  \,\,\,\, \,\,\,  \label{kop2}  \\    
+\,n \int_x \lo_x   \int_0^t d\tau\,  %
\exp{[(\lo_x + \fo_R^\prime )\,\tau]} \, \lo_x  g(x)\,  %
\exp{[-  \fo_R^\prime \,\tau]}\,  \nonumber
\end{eqnarray} 

\  Here expression $\, \lo_x  g(x)\,$ in fact acts as operator 
\[
  \lo_x \, g(x)\,=\, (\nabla_\rho\, g(x))\cdot  [\,V +T\nabla_P\,]\,\,\,, \, 
\] 
and we took into account that, obviously, 
\[
\int_x  \lo_x \, g(x)\,=\,0 \,\, 
\]

\,\,\, 

{\bf 3}.\,  Eventually we are mainly interested not in the CFs   %
themselves or their GF but in the BP's distribution function (DF)
\[
 F_0(t,R,P)\,=\, \mathcal{F}\{\psi =0\}\,=\, \mathcal{C}\{\psi =0\}\,
\]
From its viewpoint, the argument \,$\psi(x)$\, serves as  %
``thermostat random field variable'' responsible for both  %
``stochastic agitation'' of BP's velocity and its  %
``irreversible relaxation'', while the operation   %
\,$\psi(x)\rightarrow 0$\,, - if performed after all calculations, -  %
 as ensemble averaging over thermostat.
In this sense, two terms in Eqs.\ref{kop1} and \ref{kop2}  %
can be interpreted as ``random (Langevin) source'' and  %
``kinetic operator'', although, obviously, such separation  %
is not unambiguous. 

Advantage of representation Eqs.\ref{kc}-\ref{kop2},  -   %
in comparison with its origin, i.e. evolution equation  Eq.\ref{cfe_},  -   %
is in that it  makes explicit visual step from instant inter-particle    %
potential interaction to time-distributed collision events.  
This becomes quite clear if we eliminate action of   %
the thermostat  in expression (\ref{kop1}) or (\ref{kop2})  by  %
removing  \,$\psi(x)\,$ and \,$\fo$\, or  \,$\fo^\prime$\,, so that  %
exact Eq.\ref{kc} simplifies approximately to 

\begin{eqnarray}  
\ko (t) \approx \ko^\prime (t)\,\approx\,  \label{akop_}  %
n \int_x  \lo_x   \int_0^t d\tau\, \,\times     %
\,\,\,\,\,\,\, \\ \times \,\,  \nonumber  %
\exp{[(\lo_x - V\cdot\nabla_R )\,\tau\,]}\, \lo_x\,  g(x)\,  %
\exp{[\,V\cdot\nabla_R \,\tau\,]}\,  %
\approx\, \,\,\, \\ \,\approx\,  \nonumber  \,   
\frac nT\,\int_0^\infty d\tau \,\nabla_P\cdot \int_x  \Phi^\prime(\rho)   \, 
 \,\times     %
\,\,\,\,\,\,\, \\ \times \,\,  \nonumber  %
\exp{[\,\lo_x\,\tau\,]}\,  g(x)\,  \Phi^\prime(\rho)\cdot   %
(V+T\nabla_P)\, \equiv\,\widehat{B} \,\,\,  %
\end{eqnarray} 

This operator $\,\widehat{B} \,$ is  %
 nothing but usual (although non-standardly %
written \cite{p0806}) Boltzmann-Lorentz kinetic operator (BLO)  %
describing BP-gas interaction in Boltzmannian kinetics.  

\,\,\, 

{\bf 4}.\, The last simplification  in Eq.\ref{akop_}   %
neglects contribution of BP's displacements  %
during collisions to total BP's pah, which is reasonable for rarefied gas,  %
all the more in Boltzmann-Grad limit (BGL). 

However, approximation (\ref{akop_}) on the whole  neglects also much  %
more significant matter, namely, ``geometrical contest'' of particles (atoms)   %
in collisions wit given one (BP). It means merely that   realization of  %
any particular current collision is conditioned by all before happened  %
collisions:\, if one of them had different impact parameter or  %
had no place at all, then all later collisions   %
also would have different impact parameter values, moreover,  %
almost surely would be prevented at all. 
Therefore, factually, differential cross-section of   %
current collision is very  complicated function of its  pre-history.  %

In other words, differential cross-section of   %
current collision is highly irregular function of initial state of the system,. %
Moreover, so much irregular that we certainly can not    %
speak about its time-average value and, hence, its {\it \,a priori\,}  %
value. Indeed, number of initial gas parameters, which potentially  %
may influence on BP's motion during time \,$t$\,,  %
grows with time roughly   \,$\propto n\, (u_0t)^3$\, (with \,$u_0$\,  %
denoting characteristic thermal velocity) while number of  %
BP's trajectory parameters \,$\propto t/\tau_0\sim u_0t/\lambda   %
\sim n\,a^2 u_0t\,$\, (with \,$\tau_0$\, being characteristic BP's  %
free path time), that is \,$\sim (u_0t/a)^2$\, times smaller.  %
Clearly, so relatively small number of collision events in no way  %
is sufficient for their time averaging in respect to all  %
of their potential reasons. 

Consequently, there are no statistical  %
grounds to assume a priori definite (differential) cross-section  %
for them. Moreover, the deeper we go to BGL,  %
the lesser are such grounds (see also e.g. \cite{p1,tmf,ufn1}.  %
for similar argumentation). 

\,\,\, 

{\bf 5}.\,   The aforesaid is fully ignored    %
in the ``Boltzmannian kinetics'' approximation (\ref{akop_}).  %
It can be rewritten,  under same simplification (rejecting  %
BP's shift during collision),  in the form 

\begin{eqnarray}  
\ko^\prime (t)\,\approx  \,n \int_x  \lo_x \,    \label{akop}  %
\exp{[\,\lo_x\,t\,]}\, g(x)\,  \, \,, \,   
\end{eqnarray}

\  which visually claims uniform distribution   %
of collision's impact parameter (two-dimensional   %
\,$\rho $\,'s projection onto plane \,$\perp\,u$\,).  
To compare this with  %
the exact Eq.\ref{kop}, the latter  can be expressed by 

\begin{eqnarray}  
\ko^\prime (t)\,=  \int_x\, [\,n+\psi(x)]\, \lo_x \,   %
\, \times \,\,\,\,\,\,\,\, \,\, \nonumber  \\ \,\times  \,\, %
\exp{[\,\lo_x\,t\,]}\, \, \widehat{\sigma}_x\{t,\psi\}\, g(x)\,  %
\,\, ,\, \,\,\,\,\,\,\,\,\,\,\,  \,\,  \label{kop_}   \\    
\widehat{\sigma}_x\{t,\psi\}\, =\,  %
\exp{[\,-\,\lo_x\,t\,]}\,  %
\exp{[(\lo_x + \fo^\prime_R )\,t\,]}\, \exp{[\,-  %
 \fo^\prime_R \,t\,]}\,      \nonumber 
\end{eqnarray} 

\   Here operator  \,$\widehat{\sigma}_x\{t,\psi\}\,$\,  %
 (together with also important factor \,$n+\psi(x)$\, in place of \,$n$\,   %
on the left) represents now randomly non-uniform impact  %
parameter distribution. 
 
Evidently, it involves all the past evolution time, thus, potentially  %
all atoms what might achieve BP after its start, and establishes  %
some statistical connection of current collision to   %
micro-state of the rest of gas and, hence, to all earlier collisions. 

Let us show that such connection  survives,    %
moreover, remains substantial and practically important,    %
 under BGL.  

\,\,\, 

{\bf 6}.\, For this purpose, we have to return from Eq.\ref{kop_}   %
 to equivalent Eq.\ref{kop1} since it  clearly  %
distinguishes total evolution time \,$t$\, and much smaller  %
``inner time'' of collision \,$\tau$\,. 
Then, for transparent transition to BGL there,  make scale transformations 

\begin{eqnarray}  
n\,\Rightarrow\, \frac n{\xi^2}\,\,\,, \,\,\,\,\, %
\Phi(\rho)\,\Rightarrow\,  %
\Phi\left (\frac \rho\xi  \right)\,  \,\,, \,\,\, \label{st} %
\end{eqnarray} 

\  where\, \,$\xi\,\rightarrow\,0$\,, and simultaneously, in the integrals  %
over \,$x=\{\rho,p\}$\,  and \,$\tau$\,,  %
make changes of integration variables   and the ``thermostat   %
field'' variable, as  follow, 

\begin{eqnarray}  
\rho \,\Rightarrow\, \xi\,\rho\,\,\,, \,\,\,\,\, %
\psi ( \xi\rho, p)\,\Rightarrow\, \frac {\psi  ( \rho,p)}  {\xi^2}\,   %
 \,\,, \,\,\,\,\, \label{st_} %
\tau \,\Rightarrow\, \xi\,\tau \,\,\,\, \,\,\,\, %
\end{eqnarray} 

\  The middle of these changes combines scale transformation,   %
like that of the mean gas density in (\ref{st}), and   %
replacement \,$\psi(\xi\rho,p)\Rightarrow \psi(\rho,p)$\,.  
The latter should be applied, - inside \,$\fo$\, and  \,$\fo^\prime$\,   %
or similar objects, - together with functional derivative transform   

\begin{eqnarray}  
\frac {\delta}{\delta \psi ( \xi\rho, p)} \,\Rightarrow\,  %
 \frac  {\delta} {\xi^3 \,\delta \psi  ( \rho,p)}  \,\, \,\,\,\,\, \label{st__} %
\end{eqnarray} 

\  This rule reflects invariance of ``number-of-particles  operator''
in respect to our changes,    

\begin{eqnarray}  
\int_x \psi(x)\,\frac {\delta}{\delta \psi(x) }  \,\equiv\,  
\int d^3p \int d^3\rho \,\,\, \psi(\rho,p)\,  %
\frac  {\delta}{\delta \psi ( \rho, p)}\,=\,   %
\,\,\,\, \nonumber  \\ \,=\,  \nonumber 
\int d^3p \int \xi^3\, d^3\rho \,\, \,\psi(\xi \rho,p)\,  %
 \frac  {\delta} {\delta \psi  ( \xi\rho,p)}  \,\, \,
\end{eqnarray} 

Taking all this into account, it is easy to verify that  in the BGL,  %
\,$\xi\rightarrow 0$\,,  %
 both the evolution operators \,$\fo$\,   %
(\ref{fo}) and  \,$\fo^\prime$\, (\ref{fop})  are rescaled  equally as   

\begin{eqnarray} 
\fo\,,\,\fo^\prime\,\,\Rightarrow\, \, \label{lfo}    
\frac {\fo}\xi   \,=\,\frac 1\xi \int_x  [n+\psi(x)]\,   \lo_x\,     %
\frac {\delta }{\delta \psi(x)} \,\,\, , \,\, 
\end{eqnarray}

\  while both the corresponding  ``pseudo-kinetic'' operators    %
(\ref{kop}) tend to one and the same limit, 

\begin{eqnarray}  
\ko(t)\,,\,\ko^\prime (t)\,\,\Rightarrow\,  \,    \ko_\infty\,=\, 
 \int_x \psi(x)\, \lo_x\,   g(x)\,\,+\, \,\,\,\,\, \, \label{lko}  \\ \,+ 
\int_x [\,n+\psi(x)]\, \lo_x   \int_0^\infty  d\tau\,  %
\,    \times\,\,\, \nonumber \\ \,\times \,\,  %
\exp{[(\lo_x + \fo )\,\tau\,]}\, \lo_x\,  g(x)\,  %
\exp{[\,-  \fo \,\tau\,]}\, \, \, \, \,   \nonumber 
\end{eqnarray} 

At that, according to above consideration, 

\begin{eqnarray}  
F_0(t,R,P)\,\Rightarrow\,  %
\left\langle\, e^{\,(\,-V\cdot\nabla_R + \ko_\infty\,)\, t}\,   %
\right\rangle \, F_0(0,R,P)\,\,\, \,\,\,  \label{lke}  
\end{eqnarray} 

\ with angle brackets meaning statistical averaging as defined by 
\[
 \langle\, \dots\,\rangle \,=\, [\,\dots\,]_{\psi =0} \,\,\, 
\]

\   Of course, after transition to BGL, results of the averaging depend  %
on the composite  parameter \,$a^2n =(\pi\lambda)^{-1}$\,  %
as the whole only. 

In particular, on average the limit random ``pseudo-kinetic''   %
operator (\ref{lko})  coincides with the Boltzmann-Lorentz one   %
from (\ref{akop_}),  
\begin{eqnarray}  
\langle\,    \ko_\infty\,\rangle \,=\,\widehat{B}\,\,\,, \, \nonumber  
\end{eqnarray} 
while for its variance Eq.\ref{lko} formally yields 

\begin{eqnarray}  
\langle\, \ko_\infty^2\,\rangle \,-  %
\langle\, \ko_\infty\rangle ^2  \,=\,   %
n^2 \int_x \int_y  %
\int_0^\infty  d\tau  \int_0^\infty  d\tau^\prime\, \,  %
\times \,\,\,\, \,\,\,\,\,  \label{kov} \\ \times\,\,    \nonumber 
\lo_x\,[\, e^{\lo_x\,\tau}\,\lo_y\,  %
e^{(\lo_x +\lo_y)\, \tau^\prime\,} \, \lo_x\,  %
e^{-\lo_y\,(\tau +\tau^\prime\,)}\,   %
\, -\, \,\, \\ \,-\,   
e^{\lo_x\,(\tau +\tau^\prime\,)}\, \lo_x\,  %
\lo_y\, e^{-\lo_y\,\tau^\prime\,}\,]\,  \lo_y\,g(y)\,g(x)\,\,\, , \,\,  \nonumber
\end{eqnarray} 

\  where, clearly, \,$x$\, and \,$y$\, are two different atom's   %
phase points (each pointing to  momentum and   %
relative distance). 
This expression, by its nature,  %
represents just mutual interference of gas atoms  %
in possibilities of their encounters with BP. 

\,\,\, 

{\bf 7}.\,  Thus, we have demonstrated that in general BGL  %
produces a non-trivial theory  %
principally and quantitatively different from Boltzmann's   %
kinetics. 

At that, all the CFs possess also non-trivial non-zero limits,  %
though they require special careful consideration since are  %
singular functions with two ``infinitely strongly''   %
different spatial scales, \,$a$\, and \,$\lambda$\,. 

It is useful to take in mind that corresponding  %
inter-particle statistical correlations  %
have no an ``autonomous'' physical meaning (``mechanism'') but,  %
in essence, are originated   %
by mere knowledge about past BP's walk (path).  
In other words, about practically observed rate of   %
system's evolution to final statistical   %
equilibrium where \,$\nabla_R F_0 \rightarrow 0$\,. 
Indeed, a greater value of BP's path gives evidence of  its faster  %
diffusion and lesser rate (relative frequency) of its collisions,  %
or smaller efficiency (effective cross-section) of collisions,  - and vice versa, -  %
which is just the source of BP-atoms correlations. 

From formal viewpoint of  BBGKY equations,  %
inequalty \,$\nabla_R F_0 \neq 0$\,  %
implies violation of equilibrium (Maxwellian) character  %
of BP's velocity distribution, which in turn inevitably    %
induces various inter-paticle correlations. 

From physical viewpoint, the system (gas) forgives    %
deviations of rate (relative frequency  %
and efficiency) of BP's collisions   %
from its imaginary ``mean value''. Therefore, each particular  %
realization of BP's life (trajectory) randomly acquires its own   %
unique time-averaged collision rate \cite{i1,i2,last,bk3,bk12,ufn1}. 
Then all atoms somehow (actually or virtually)   %
 involved into BP's life become ``guilty of''   %
(correlated with) its unpredictable result.

\subsection{Why alternative collision boundary  %
conditions seem good}

Advantage of our probability-theoretical formulation of the hard-ball  %
collision rules, - i.e. conditions (\ref{cr}) and (\ref{crs}) or (\ref{rc}), -   %
is that it  %
creates no questions when being applied to configurations   %
characterized by  %
two or several \,$|\rho_{j}|\rightarrow a$\, at once. 
The matter is that different conditions from a set (\ref{crs}) are freely  %
compatible one with another, because all they are mutually  %
commutative, in contrast to conventional conditions from (\ref{ccrs}). 

Due to this fact, now we obviate the necessity  %
of artificial division of many-particle configurations  %
like three- or multi-particle  %
collisions into almost simultaneous pair collisions. 
At the same time, we can consider arbitrary chains  %
and graphs of more or less close pair collisions, being guaranteed  %
for smooth unambiguous transition from them   %
to ``many-particle collisions''. 

Thus, the related formalism, in contrast to conventional one,   %
is logically complete  %
and unambiguous. This is good stimulus for investigation  of   %
its principal consequences and practical utility. %
At present, we  confine ourselves by its preliminary discussion only. 


\section{Principal properties of  %
hard-ball BBGKY hierarchy and expected solutions to it}

\subsection{Possibility of continuation to the whole \,$\rho$\,-space}

{\bf 1}.\, According to above derivation of our of hard-ball  %
BBGKY hierarchy, Eqs.\ref{qn},  %
the functions \,$Q_k$, (with $k>0$) there are such  %
that can be continuously extended into (physically forbidden)  %
regions \,$|\rho_j|<a$\,, for some or each of \,$1\leq j\leq k$\,. 
In corresponding version of the theory,   %
the collision boundary conditions (CBC)  %
(\ref{crs}) must be replaced by ones which follow  %
from Eq.\ref{req_}, that is 

\begin{equation}
\Omega_j\cdot (\nabla_{p_j} -\nabla_P)\, Q_k(\cdot)\,=  %
\,0\,  \,\,\,\,  \texttt{at}\,\,\,\,\, %
|\rho_j| \,\leq\, a\,\,\,    \label{bcs}
\end{equation}

At that, besides, the Eqs.\ref{qn} themselves must be  %
modified too, since, - in view of Eq.\ref{req}, - the limit expressions  

\[
\frac {\Phi^\prime (\rho_j)\cdot (\nabla_{p_j} -\nabla_P)\,Q_k}  %
  {Q_k}\, \Rightarrow \, \gamma_k(\cdot)\,\,
\]
  
\ may be thought having non-zero finite values,   %
in spite of zero in Eq.\ref{req_}.  %
Therefore, now we should write 

\begin{eqnarray}
\dot{Q}_k\,=\,\gamma_k\,Q_k\,-V\cdot\nabla_R\, Q_k + %
\sum_{j=1}^k\,(V-v_j)\cdot\nabla_{\rho_j}\, Q_k\, +  %
\nonumber  \\ +\,   
\,na^2 \oint \!\int dp_{k+1}\,\, (\Omega\cdot(v_{k+1}-V)) \,  %
\times\, \nonumber \\ \times \,
G_m(p_{k+1})\,\, Q_{k+1}(\rho_{k+1}=a\Omega)\,\,\,  %
\,\,\,\,\,\,\,\,\,\,\,\,\,  \label{iqn}
\end{eqnarray}
 
\ Here \,$\rho_j$\, take arbitrary values,\, and the  %
``sources'' \,$\gamma_k\,Q_k$\, have appeared, which can differ from zero in the %
forbidden regions only (\,$\gamma_k(\cdot)\neq 0$\, if and only if $|\rho_j|<a$\,  %
for at least one of \,$1\leq j\leq k$)\, and must be chosen     %
in such a way that all the conditions (\ref{bcs})  %
are satisfied. 

Possible advantage of this formalism, as compared with Eqs.\ref{qn}  %
plus \ref{crs}, is obvious:\, it allows to represent all \,$Q_k$\,'s  %
by evident iteration series free of CBC which now   %
transform into integral equations for the sources. 

\,\,\, 

{\bf 2}.\,  In the conventional approach too there is  %
possibility to go to the whole \,$\rho$\,-space (``without holes'')   %
if introducing proper source terms.  %
In particular, sources can be concentrated  at boundaries of %
forbidden regions, i.e. at \,$|\rho_j|=a$\,.  %
In such way one can include all the CBC (\ref{ccrs}) to  %
sources and thus automatize their use. 

The corresponding form of BBGKY equations was mentioned e.g.  %
in \cite{pg} as ``pseudo-Liouville''.  %
In principle, it brings help for objective evaluation of  %
contributions from specific ``unpleasant''  %
many-particle configurations and events, avoiding their  %
artificial sorting out and thus preventing crucial losses  %
of standard formalism \cite{cpg,pg}.  %
Such ability of ``pseudo-Liouville'' representation  %
was demonstrated in \cite{hs1}.

\subsection{Density of collisions drifts with   %
centre-of-mass velocity}

{\bf 1}.\, The integral terms in Eqs.\ref{qn} (as well as in Eqs.\ref{fn}), -  %
responsible for collisions with ``outside'' particles, -  %
are determined by space-angle averaging in \,$\rho$\,-space %
and vector functions like

\begin{eqnarray}
A_{k\,j}\,=\,\oint \Omega\, Q_{k}(\rho_{j}=a\Omega)\, \, \label{ak}
\end{eqnarray}

\  The same functions appear when averaging our conditions 
(\ref{crs}) which yields

\begin{eqnarray} 
(\,\nabla_{p_j}-\nabla_P\,)\cdot A_{k\,j}\,=\,0\,\,\,  \label{acr}
\end{eqnarray}

In particular, at $k=1$ the latter formula can be rewritten as 

\begin{eqnarray} 
\nabla_{u_1}\cdot A_{1\,1}\,=\,0\,\,\, \nonumber   
\end{eqnarray}

\ This equality shows, first, that \,$A_{1\,1}$\, has meaning  %
of flow of two-particle (BP-atom) correlations in %
space of relative velocity \,$u_1=v_1-V$\,.  
Second, this flow is either purely rotational or simply  %
constant vector. Following the ``Ockham razor'' principle, 
it is natural to choose the latter variant,  %
since the former one has no visible physical sense (at least  %
in theory where particles do not rotate).  
Then

\begin{eqnarray}
A_{1\,1}\,=\,\oint \Omega\, Q_1(\rho_1=a\Omega)\,=\,  %
A_{1\,1}(t,R,P+p)\, \,\,, \nonumber 
\end{eqnarray}
\ that is the space angle integral in collision term of the first of Eqs.\ref{qn},  %
for \,$F_0$\,, depends on total momentum of colliding pair only.

\,\,\,

{\bf 2},\,  Similar statement can be concluded  %
as true in respect to the function \,$Q_1(\cdot)$\, itself, 
which characterizes local (\,$R$\,-dependent) rate  of (various types of)  %
collisions. Namely, 

\begin{eqnarray}
Q_1(\rho_1=a\Omega)\,=\,Q_1(t,R,\Omega,\,P+p)\,\,\, \label{q1}
\end{eqnarray}

\ In other words, rate of collisions depends  %
on centre of mass velocity, \,$(MV+mv)/(M+m)$\,  (CMV) only,  %
but not  separately on BP's and atom's velocities. 

Consequently, CMV is characteristic velocity of drift of  (mean)    %
collision rate in real  configurational space. 

\,\,\,

{\bf 3}.\,  Analogous discussion of functions  
 
 \begin{eqnarray}
B_{k}\,=\,Q_{k}(\rho_1=a\Omega_1,\,\dots\,, \rho_k=a\Omega_k)\, \, \label{bk}
\end{eqnarray} 

\  and functions\, \,$\oint_1\!\dots\!\oint_k\, \Omega_1\dots\Omega_k\,B_k\,$\,  %
(multi-vector, or tensor-like, objects) pushes to suppose that  %
all they depend on momenta or velocities  %
through single variable, \,$P+p_1+\dots +p_k$\, or, equivalently,     %
\,$(k+1)$\,-particle CMV\, %
  \,$(MV+mv_1+\dots +mv_k)/(M+km)$\,. In particular, 
  
\begin{eqnarray}
B_k\,=\,B_k(t,R,\Omega_1 ,\,\dots\,,\Omega_k,\,P+p_1+\dots + p_k)\,\,\, \label{qk}
\end{eqnarray}

That are statistical characteristics of   %
randomness, or fluctuations, of collisions' rate.  
Although we can interpret them also as characteristics of local   %
rates of specific many-particle configurations and events.   %
Anyway, taking in mind that the particles   %
might belong to either one and the same ``coherent''  %
process or to two or several  competitive or concurrent processes.  
Property (\ref{qk}) allows to expect that  %
these characteristics  drift with CMVs  \,$(MV+mv_1+\dots +mv_k)/(M+km)$\,. 
 
\,\,\,

{\bf 4}.\,  These reasonings  are  %
in agreement with results of our first  %
analysis of BBGKY hierarchy in \cite{i1}     %
in the framework of ``collisional as approximation''  %
suggested there (see also \cite{i2,p1,p1007}). 
Simply, relative motion of colliding   %
(or close) particles is inner constitutient part of  %
their collision (or encounter)  and therefore   %
does not contribute to spatial drift  %
of collision (encounter) as the whole   %
which thus acquires centre of mass velocity (CMV). 

\,\,\, 

In the rest of this paper, let us point out how   %
solutions to BBGKY equations  %
are constructed, and why they %
possess properties like  (\ref{bk})-(\ref{qk}), thus cardinally    %
destroying naive Boltzmann's molecular chaos,  %
by transmuting its relaxation rates into random quantities.
As before, we shall exploit particular example of ``BP in ideal gas'',  %
but our reasonings will be quite general. 


\section{Characteristic structure of solutions to (hard-ball)  %
BBGKY hierarchy}

\subsection{Stationary solutions and Liouville operators}

{\bf 1}.\, First, discuss stationary solutions of Eqs.\ref{qn} or,   %
equivalently Eqs.\ref{cn} (or Eq.\ref{gqe}), when all \,$\dot{Q}_k=   %
 \dot{C}_k=0$\,.   
Clearly, such solutions  must be at once  spatially homogeneous:  %
\, \,$\nabla_R Q_k=\nabla_R C_k=0$\,. Designating them by  %
  \,$Q_k^o =Q_k^o(\rho_1,\,\dots\,P, p_1\,\dots\,)$\,  and  %
 \,$C_k^o$\,, we thus have   

\begin{eqnarray}
0\,=\,\sum_{j=1}^k\,\widehat{L}_j\, Q_k^o\, +  %
\label{qn0} \\ +\,   
\,na^2 \oint \!\int dp_{k+1}\,\, (\Omega\cdot(v_{k+1}-V)) \,  %
\times\, \nonumber \\ \times \,
G_m(p_{k+1})\,\, Q_{k+1}^o(\rho_{k+1}=a\Omega)\,  %
\,\,\,\, \nonumber
\end{eqnarray}

\  and similarly for \,$C_k^o$\,, where  %
\,\,$\widehat{L}_j\,$ means partial Liouville operator  %
(\ref{lj}) in the hard-ball limit defined by expression  %
 \,$\widehat{L}_j=(V-v_j)\cdot\nabla_{\rho_j}\,$  for $\,|\rho_j|>a$\,  %
and by the CBC at $\,|\rho_j|=a$\,, (\ref{ccrs})  %
or (\ref{crs}) for \,$Q_k^o$\,'s   %
and (\ref{crc}) or (\ref{rc}) for \,$C_k^o$\,'s. %
One may supply these equations with some boundary conditions  %
at infinity, e.g. \,$Q_{k}^o(\rho_{j}=\infty)\,=\,Q_{k-1}^o$\,  %
or \,$C_{k}^o(\rho_{j}=\infty)\,=\,0$\,.  %

{\bf 2}.\, Of course, there is trivial  %
solution \,$Q_k^o=$\,const\,, at \,const\,$=1$\, describing (standardly normalized)  %
canonical thermodynamically equilibrium state of the system,   %
free of any information about BP's position and inter-particle  %
statistical correlations. 
Notice that it appears from exact  %
time-dependent solution of Eqs.\ref{qn}, with  %
 initial conditions (\ref{ic_}), when removing its dependence on  %
BP's position\,$R$\, by integration over \,$R$\,:

\begin{equation}
\int Q_k \,dR\,=1\,  \,\, , \,     \label{iq0}
\end{equation}

\ since \,$\int Q_k(t=0)\, dR\,=\,1$\,. 
Thus, all mathematical problems induced by Eqs.\ref{qn} can  %
be killed by the single integration. 

We, however, are interested in essence of these problems  %
and, therefore, in question whether there are non-trivial  %
(non-constant) solutions of Eqs.\ref{qn0}, i.e. such solutions that,  %
in terms of related cumulant functions (CF) (\ref{hqc}),   %
 \,$C_k^o\neq 0$\, at \,$\rho_j\neq\infty$\,  %
and \,$C_k^o\rightarrow 0$\, when at least one \,$\rho_j\rightarrow\infty$\,.  
Our principal statement is that such solutions indeed exist.  %

\,\,\, 

The point is that any solution to Eqs.\ref{qn0} is nothing but chain   %
of \,$k$\,-dimensional projections of a solution to  infinitely  %
many-dimensional Liouville equation 

\begin{equation}
[\,-\,V\cdot\nabla_R\,+ \sum_{j=1}^N\, \widehat{L}_j\,]\,  %
Q_N^o \,=\,0\,\,\,,\,\,\,   \label{il}
\end{equation}

\ with large \,$N\Rightarrow\infty$\,. 

Undoubtedly, at any \,$N$\,  %
this equation has infinitely rich variety of   %
non-constant solutions, both dependent and independent on \,$R$\,.  %
Their peculiarity is that any of them  %
keeps constant (invariant in respect to translations) along  %
phase trajectories of the system, in its \,$(N+1)$\,-particle phase space,   %
but can change along different transversal directions, - i.e. from one  %
trajectory to another, - in  arbitrarily irregular way. 

For example, we may establish it to be non-zero at such  %
trajectories only on which BP's collisions with several or even all of \,$N$\,  %
atoms definitely take place (sometime or, may be, within a given time interval). 
Arbitrary linear or nonlinear  %
combination of such or otherwise specified functions  %
of \,$(N+1)$\,-particle phase point  %
belongs to the same class of functions, i.e.  %
eigenfunctions of \,$(N+1)$\,-particle Liouville operator with   %
zero eigenvalue.  %

By these means, it is possible to compose \,$Q_N^o$\, producing,  %
under \,$N\rightarrow\infty$\,,  %
a sequence of standardly connected  %
partial functions \,$Q_k^o$\, (\,$k=0,1,2,\dots$\,)  %
representing  non-trivial solution to Eqs.\ref{qn0}. 

By their construction, such \,$Q_k^o$\, (\,$k=1,2,3,\,\dots$\,)  %
must look like {\it \,localized  excitations\,}  %
of correlation field \cite{fn3}. At that, since they are stationary,  %
they possess definite symmetry in respect to time reversal,  %
that is equally include post-collision and pre-collision  %
inter-particle correlations.

Besides, due to high degree of arbitrariness in such  %
constructions,  their result \,$Q_k^o$\,'s satisfying Eqs.\ref{qn0}  %
simultaneously can be furnished with an infinite set of free parameters. 

\,\,\, 

{\bf 3}.\, Spatial extension, in \,$\rho_j$\,-spaces,  %
of the localized excitations is determined by the ``collision  %
integrals'' in Eqs.\ref{qn0}. Naturally, this extension, let be denoted  %
by \,$d$\,, is of order of %
characteristic BP's ``mean free path'' \,$\sim\lambda =(\pi a^2n)^{-1}$\,.  
More precisely, it may depend also on BP-atom relative velocities, since,  %
strictly speaking, this integrals determine sooner characteristic  %
``rate'' of BP's collisions, \,$\sim u_0/\lambda$\,, - with   %
 \,$u_0=\sqrt{T/m^\prime}$\,,\, \,$m^\prime=(mM/(m+M)$\,,, -   %
so that 
\begin{equation}
d\,\sim\, \lambda\,|v-V|/u_0\,\,\,\,\,   \nonumber  
\end{equation}
Then at \,$|\rho_j| \lesssim d $\, one can neglect effect of collisions with  %
outer particles and write approximately  

\begin{equation}
\sum_{j=1}^k\, \widehat{L}_j\,  %
Q_k^o \,=\,0\,\,\, \,\,   \label{kl}
\end{equation}

\ (\,$k>0$), instead of (\ref{qn0}), with contact boundary values  %
 \,$Q_k^o(|\rho_j|=a)$\, considered like ``initial conditions''  %
to \,$(k+1)\,$-particle phase trajectories. 
Solutions to them as well allow for  %
many free parameters. In particular, total \,$(k+1)$\,-particle %
momentum \,$P_k=P+\sum_{j=1}^k p_j$\, can be treated as one of parameters,  %
because it is conserved by the Liouville operator there.

\subsection{Quasi-stationary asymptotics of inter-particle correlations}

{\bf 1}.\, Another important statement to argue is that  %
inter-particle correlations, which are induced by BP's walk after start from  %
non-correlated state (\ref{ic_}) and described by Eqs.\ref{qn}   %
or Eqs.\ref{cn}, do not disappear with time, instead going to  %
approximately stationary asymptotics satisfying Eqs.\ref{qn0}. 

This is clear already from the first of DVR (\ref{qcdvr})  %
if rewriting it as  

\begin{eqnarray}
\frac {\partial \ln\, Q_0 }{\partial \ln\, n}\,= %
n\int_{|\rho_{1}|>a} \!\int_{p_{1}} G_m(p_{1})\,  %
\left[\,\frac {Q_{1}}{Q_0} - 1\,\right]\,\,\, \, \label{dvr1} %
\end{eqnarray} 
 
\ and taking into account that all \,$Q_k$\,s decrease with time  %
proportionally to \,$Q_0$\,, -  \,$Q_k\propto Q_0$\,, -  %
while \,$Q_0$\, decreases with (long enough) time by a  
``diffusive law'', 

\begin{eqnarray}
Q_0(t,R,P\,|\,n)\,\sim\, (4\pi Dt)^{-3/2}\,  %
\Psi(R^2/4Dt)\,\,\,\, \, \label{dl} %
\end{eqnarray} 

\ Here \,$D=D(n)$\, is characteristic diffusivity,\, and function   %
 \,$\Psi(\cdot)\geq 0$\, (\,$\Psi(0)=1$\,) and generally is not  %
exponential (reducing to exponential at \,$M/m\rightarrow\infty$\, only)  
\cite{p1,p0710,p0802,p0803,tmf,p1007,ufn1,last}. For dilute gas,  %
or under BGL    %
\,$D(n)\sim \lambda\sqrt{T/M} = \sqrt{T/M} /(\pi a^2n) \propto 1/n$\,. 
From Eqs.\ref{dvr1} and \ref{dl} it follows that at large time  %
(``far at kinetic stage'') both sides in Eq.\ref{dvr1} become a function  %
of single dimensionless  argument \,$\zeta =R/\sqrt{2Dt}$\,,   %
that is has non-zero finite limit when  \,$t\rightarrow\infty$\, at any 
fixed $\,\zeta$\, or \,$R\,$ . 

This means, in turn, that ratios \,$Q_k/Q_0$\, have non-zero  %
long-time limits, moreover, quantities   %
that factors  \,$Q_1/Q_0 -1$\, and  %
 \,$Q_{k+1}/Q_k -1$\,, - representing BP-atoms correlations,  %
stay non-zero. 

\,\,\, 

{\bf 2}.\, Hence, we can identify (accurate to some common positive multiplier)  %
the limit ratios \,$Q_k/Q_0$\,, on one hand, and the above  %
suggested functions \,$Q_k^o$\,  %
satisfying stationary Eqs.\ref{qn0}, on the other hand. At that,  \,$Q_k^o$\  %
acquire  argument \,$R$\,, or \,$\zeta =R/\sqrt{2Dt}$\,,  %
in the role of free parameter, among other possible ones. 

Correspondingly, we can use Eqs.\ref{qn0} or  %
Eqs.\ref{kl} as a tool for constructing qualitatively correct  %
approximations of actual non-stationary solutions to Eqs.\ref{qn}.

\subsection{Collisional approximation} 

{\bf 1}.\, Basing on all the aforesaid, let us try to separate  %
``fine details'' of inter-particle correlations, - at small relative distances  %
  from \,$|\rho| \sim a$\, up to \,$|\rho| \sim\lambda$\,,    %
- and overall value of these correlations varying at larger  %
distances  \,$|\rho| \gtrsim\lambda$\,, - by assuming that the first  %
take a constant shape at kinetic stage of evolution,   %
while the second continues to  %
change together with probability distribution of BP's position. 
The first are correlations s inside clusters   %
of close particles    %
conserving their summary momentum \,$P_k=P+\sum_{j=1}^k\,p_j$\,.  %
The second is represented by a set of mean densities of   %
\,$(k+1)$\,-particle clusters  in real configurational space. 

Naturally, these densities are drifting with the  %
centre-of-mass velocities (CMV) \,$V_k=P_k/M_k\,=$ %
  \,$V+(m/M_k)\sum_{j=1}^k\,u_j$\, (where \,$M_k=M+km$\,) and   %
therefore must be ``attached'' to centre-of-mass position  
\begin{equation}
R_k\,=(MR+m\sum_{j=1}^k\,r_j)/M_k\,=  %
 \,R+ \frac m{M_k}\sum_{j=1}^k\,\rho_j\,\, \label{rk} 
\end{equation} 
 
To separate the latter  %
from relative motion of particles inside clusters,   %
let us rewrite Eqs.\ref{qn} in the form  %

\begin{eqnarray}
\dot{Q}_k\,=\,-V_k\cdot\nabla_R\, Q_k \,+\,\,\,\,\,\,\, \,\,\,\, \label{mqn}  
\\ \,+\,  
\sum_{j=1}^k\,[\,\widehat{L}_j + \,\frac m{M_k}\,  %
u_j\cdot \nabla_R\,]\,Q_k\, +\, \nonumber 
\\ \,+\, 
na^2 \oint \Omega\cdot \langle\, (v -V)\,  %
Q_{k+1}(\rho_{k+1}=a\Omega,\,p_{k+1}=p)\,\rangle_p\,\,\,,\, \nonumber 
\end{eqnarray}

\  where \,$p=mv$\, and angle brackets denote averaging over equilibrium  %
atom's momentum distribution: 

\begin{eqnarray}
\langle\,\dots\,\rangle_p\,=\,  %
\int \dots\, G_m(p)\, dp\, \,\, \nonumber 
\end{eqnarray}

We thus excluded parts \,$(m/M_k)\,u_j\cdot \nabla_R Q_k\,$  %
from the first right-hand term and added them to \,$\widehat{L}_j$\,.  %
Due to this transfer, evolution operators defined  %
in first and second rows  %
of Eqs.\ref{mqn} now commute one with another, even in respect to  %
spatially inhomogeneous \,$Q_k$\,s, i.e. somehow depending on \,$R$\,  %
(or, equivalently, on \,$R_k$\,s). 

\,\,\,

{\bf 2}.\, Hence, we can associate these operators with evolutions of  %
``overall'' density of correlated many-particle clusters and  %
``fine'' distribution of correlations inside clusters, respectively.   %
Then, - following the claimed course, - assume that  %
at kinetic, or ``quasi-stationary'', stage of evolution,  %
for relatively close particles, approximately

\begin{eqnarray}
\sum_{j=1}^k\,[\,\widehat{L}_j + \,\frac m{M_k}\,  %
u_j\cdot \nabla_R\,]\,Q_k\, =\,0\,\,\, \, \label{mkl}  
\end{eqnarray}

\ By essence, this is the same equation as Eq.\ref{kl}, with those difference  %
only that, evidently, solutions to Eq.\ref{mkl} possess  \,$P_k$\,  %
and \,$R_k$\, among their free parameters, instead of \,$P_k$\,  %
and \,$R$\, for solutions of Eq.\ref{kl}. 

Probability-theoretical meaning of Eq.\ref{mkl}, as well as Eq.\ref{kl},  %
 is very simple:\, it states that, in statistical ensemble under  %
consideration, various sequential stages (time sections) of one and the same  %
two-particle collision, or more complex many-particle event,  %
 are represented with equal probabilities  %
(probability densities). Clearly, this is quite necessary   %
condition (ansatz) since otherwise one could not treat a given   %
configuration as time section (instant view) of    %
definite coherent collision or event as the whole.  %

 That is why thus arising  approach, for the first  %
introduced in \cite{i1}, later in \cite{i2,p1,p0806,p1007} was  %
named ``collisional approximation''. At that, generally,  %
``collisions'' are meant in wide sense as   %
chains or packs of connected  %
or competitive (``mutually interfering'', ``virtual'', etc.)   %
or merely close pair collisions, or may be  %
even ``encounters'' of   %
particles without substantial interaction. 

\,\,\, 

{\bf 3}.\,    In essence,  Eqs.\ref{kl}  %
and \ref{mkl} serve as direct analogues of the ``extended CBC''   %
(\ref{accr})-(\ref{dccr}).
To see this, let us apply similar reasonings to events  %
(``multi-particle collisions'') associated with specific   %
configurations with all \,$|\rho_j|\rightarrow a$\,. 
Considering them in the centre-of-mass frame,  %
we the have, instead of (\ref{accr}), 

\begin{eqnarray}
F_k\left( R=R_0  + \,\frac m{M_k}\sum\,u_j\,dt\,,  %
\rho =a\Omega - u\,dt,\,u\right)\,=\, \,\,  %
\,\,\,\,  \,\,\,\,\,\,\, \label{accr_}  \\ \,=\, \nonumber 
F_k\left( R=R_0  - \,\frac m{M_k}\sum\,u_j^*\,dt^*\,,  %
\rho =a\Omega + u^*\,dt^*,\,u^* \right)\,\, 
\end{eqnarray}
\  Here \,$\rho\,$, \,$\Omega\,$ and \,$u$\, replace  %
full sets of variables,  \,$dt>0$\, and   %
\,$dt^*>0$\, again are arbitrary infinitesimal quantities, and 
\[
 R_0\,\equiv\, R_k - \frac m{M_k}\sum\, a\,\Omega_j \,\,\,  
\] 
 
This condition yields, instead of (\ref{dccr}), 

\begin{equation}
\left[\,-\sum\,  u_j\cdot\left(\nabla_{\rho_j} -\,  %
\nabla_R \right)\, F_k\,\right]_{|\rho| =a}\,=\,0\,\,\,, \,\,  \label{dccr_}   %
\end{equation}

\  which coincides with Eq.\ref{mkl} as applied to vicinity  %
of multiple collision boundary, \,$\rho_j\rightarrow a\Omega_j$\,. 
Such short-cut version of Eq.\ref{mkl}, however,  %
is sufficient for deducing the collisional approximation. 

\,\,\, 

{\bf 4}.\,  Since in this approximation   %
 we neglect details of distributions of  %
inter-particle correlations in \,$x=\{\rho,p\}$\,-spaces, - %
when excluding second row of Eqs.\ref{mqn}, - we must correspondingly  %
roughen also third row there, by excluding from it now inaccessible %
``fine'' information about momenta and space-angles'  %
dependencies of boundary values of DFs or CFs.  
More precisely, information about pre-collision   %
correlations between momenta of %
actually colliding particles (BP and ``outer'' atom). 

To perform this simplification, firstly, let us apply the conventional  %
CBC (\ref{ccrs}) to express, as usually,  %
 the ``collision integrals'' in Eqs.\ref{mqn}  %
through pre-collision states. Secondly, make there replacement 

\begin{eqnarray}
Q_{k+1}(\rho_{k+1}=a\Omega,\,p_{k+1}=p)\,\Rightarrow\,   %
\,\,\, \nonumber \\ \,\Rightarrow\, %
\langle\, Q_{k+1}(\rho_{k+1}=a\Omega,\,p_{k+1}=p)\,\rangle_p\,\,  %
 \nonumber \\ 
\,\texttt{at}\,\,\,\,\,\, (\Omega\cdot (v-V))< 0\,\,\, \label{repl} 
\end{eqnarray}

\ We thus treat the ``outer'' (\,$(k+1)$\,-th) atom like ``thermostat atom'' 
whose random momentum just before its collision with BP obeys  %
purely Maxwell distribution.  %
Simultaneously, of course, we have to ignore a questionable ``fine'' dependence %
of \,$Q_{k+1}(\rho_{k+1}=a\Omega,\,p_{k+1}=p)\,$ on \,$\Omega$\,  %
at \,$(\Omega\cdot(v-V))<0$\,. It means that we prescribe uniform  %
distribution of  the collision's  impact parameter  %
\,$a\,[\Omega-u\,(u\cdot\Omega)/u^2]\,\perp\, u$\,.  %
This ansatz, however, has a little in common with   %
Boltzmamm's ``molecular chaos'' hypothesis,  %
since in general ratio \,$Q_{k+1}/Q_k$\,, as well as   %
\,$\langle Q_{k+1}\rangle_p/Q_k$\,, is different from unit and  %
possesses significant dependence on momenta of all other \,$k$\, atoms. 

After all that, one comes to equations 

\begin{eqnarray}
\dot{Q}_k\,=\,-V_k\cdot\nabla_R\, Q_k \,+\, \label{caqn}  
\widehat{B}^\dagger\, \langle\,Q_{k+1}\,\rangle_p\,\,\,, \,  
\end{eqnarray}

\ where\, \,$Q_k\,=\,Q_k(t,R,P,p_1\,\dots\,p_k)$\,, 
\begin{eqnarray}
\langle\,Q_{k+1}\,\rangle_p\,=\, \,\,\,\,\, \, \,\,\label{proj} \\ \,=\,   %
\int  Q_{k+1}(t,R,P,p_1\,\dots\,p_k,\,p)\, G_m(p)\, 
dp\,\,\,, \,\,  \nonumber  
\end{eqnarray}
\ and\, \,$\widehat{B}^\dagger =\widehat{B}^\dagger(V,\nabla_P) \,$ \,is  %
conjugated (transposed) Boltzmann-Lorentz operator defined by  %
\begin{eqnarray}
\widehat{B}^\dagger\,Q(P)\,=\,na^2  %
\int_p \oint_{(\Omega\cdot (v-V)) <0}\,  
(\Omega\cdot (v-V))\, \times \,\nonumber \\ 
\times\,\,G_m(p)\, [\,Q(P) - Q(P^*)\,]\,  %
\,\, \,  \,\,\,\, \label{cbl}
\end{eqnarray}

\ The latter is connected to the usual Boltzmann-Lorentz operator  %
 \,$\widehat{B}\,$ by operator-valued equality 
\[
\widehat{B}\,G_M(P)\,=\, G_M(P)\, \widehat{B}^\dagger\,\, 
\]

Notice that at \,$k>0$\, in Eqs.\ref{caqn} the BP's coordinate vector \,$R$\,  %
in fact plays as the centre-of-mass coordinates \,$R_k$\,,   %
since \,$\nabla_R R_k =1$\, (besides, \,$R$\, practically coincides  %
with \,$R_k$\, at \,$|\rho_j|$\,s comparable with \,$a$\,).

It should be noticed also that the same shortened  Eqs.\ref{caqn}  %
can be easy derived directly from Eqs.\ref{qn} if considering the  %
boundary DFs  (\ref{bk}) and  applying the CBC   %
extension in the form e.g. of Eqs.\ref{dccr_}.

\,\,\, 

{\bf 5}.\, The Eqs.\ref{caqn} are equivalent to equations originally deduced   %
in \cite{i1}. In spite of presence of the Boltzmann-Lorentz operator  %
in these equations, they predict crucial violation of Boltzmamm's  %
``molecular chaos'', so that certainly 
\[
 \langle\,Q_{k+1}\,\rangle_p\,\neq\, Q_k\,\,\,
\]
\ In particular, quantity \,$\langle\,Q_{1}\,\rangle_p\,$ there, -  %
which represents  %
local (space-time dependent) ensemble-averaged density, or  %
probability density, of BP-atom collisions, - does not reduce   %
to quantity \,$Q_0$\, representing BP's probability density  %
distribution  \,$F_0=G_M(P)\,Q_0$\,. 
Instead, both they are determined by all the infinite rest  %
of hierarchy Eqs.\ref{caqn}. 

Physically, this means that dynamical system under our consideration  %
possesses no {\it \,a priori\,} predictable ``probabilities of  %
collisions'' which would be same for ``almost all'' realizations of  %
the system's dynamical evolution (experiments). 
Instead, almost all experiments show their own unique {\it \,a posteriori\,}  %
``probabilities'' (relative frequencies). 

Mathematically, all that is caused by the drift terms in Eqs.\ref{caqn},  %
which state that density distributions of different sorts of collisions  %
(and many-particle events) shift in space with different  %
centre-of-mass  drift  %
velocities. Just this is formal source of inter-particle correlations.   %
It shows that correlations  %
arise in spatially inhomogeneous statistical ensembles  %
and, hence, by their nature are spatial correlations. 
On the other hand, this source appears in the foreground  %
like the ``Cheshire Cat's smile'' while the ``Cat himself'', that is %
detail microscopic background picture of the  correlations,  %
becomes invisible under the ``collisional approximation''. 

\,\,\, 

Some possibilities of this approach were presented  %
in \cite{i1,i2,p1,p1007}. It qualitatively reveals true statistics of %
``molecular Brownian motion'' and even gives its reasonable  %
semi-quantitative estimates, in particular, for accompanying  %
diffusivity/mobility 1/f-noise. 

But complete structure of solutions to basic exact BBGKY hierarchies  %
still requires serious mathematical investigation.   %
To end this paper, let us shortly discuss some of related questions.


\section{Beyond the collisional approximation}  %

\subsection{Localization of inter-particle correlations  %
and space-angle averaging in the Boltzmann-Grad limit}

{\bf 1}.\, When considering \,$\rho_j$\,-dependencies of solutions to  %
Eqs.\ref{qn} or Eqs.\ref{cn} or Eqs.\ref{qn0}, etc., it seems natural to  %
use spherical coordinates, e.g. in terms of variables 

\[
 q_j\,=\,\frac {|\rho_j|}a \,\,\,, \,\,\,\,\, %
 \Omega_j\,=\,\frac {\rho_j}{|\rho_j|}\,\,\,, \, 
\]

and transformations like 

\[
 Q(\dots\,\rho_j\,\dots)\,\Rightarrow\,  %
\oint Q(\dots\,aq_j\Omega\,\dots)\,f(\Omega)\,\,
\]

\ with proper space angle functions \,$f(\Omega)$\,, e.g.  in order to  %
extract various ``multi-pole components'' of \,$\rho_j$\,s dependent fields. 

 Motivation for such manipulation is obvious:\, the collision integrals  %
are determined by ``dipole'' components at \,$q_{k+1}=1$\,,  %
while the dynamical virial relations (DVR) involve ``scalar'' components %
integrated over all \,$q_{k+1}$\,'s values. For example, Eq.\ref{dvr1},  %
if rewritten, once more, via the cumulant functions, as 

\begin{eqnarray}
\frac {\partial C_0 }{\partial \ln{a^2 n}}\,= %
a^3 n \int_1^\infty dq_1\,q_1^2  \int_{p_1} G_m(p_1)  %
\oint C_{1}\,\,\, \, \,\, \label{dvr11} %
\end{eqnarray} 

 \,\,\,
 
 {\bf 2}.\,  The latter formula clearly prompts that  under the BGL,   %
 when \,$a^3n =a/\pi\lambda \rightarrow 0$\,  %
 (with fixed characteristic free path length   %
 \,$\lambda =(\pi a^2n)^{-1}=$\,const\,),   %
 the scalar component of \,$C_1$\, behaves like 
 
 \begin{eqnarray}
C_1^0\,\equiv\,  %
\frac 1{4\pi} \oint C_{1}\,   \Rightarrow\,   %
\frac  { S_1^0 (aq_1/\lambda)}{q_1^2}\,,=\,  %
\frac {a^2 S_1^0(|\rho_1|/\lambda)}{|\rho_1|^2}\,  \,\,\,\, \, \,\, \label{dvr1l} %
\end{eqnarray} 

\  with some integrable function \,$S_1^0(\cdot)$\,.  %
Otherwise right-hand side in (\ref{dvr11})  %
would have either zero or infinite limit.  %

Such characteristic law of localization of inter-particle (BP-atom here)  %
correlations, as in (\ref{dvr1l}), can be argued, heuristically  %
or formally, in several different ways \cite{p0710,p1105,p1203}.  

Thus, at short enough relative distancies (comparable with \,$a$\,)   %
space angle-averaged inter-particle correlation decreases inversely  %
proportionally to area \,$4\pi q^2$\, of surrounding sphere  %
(similarly to light intensity around point emitter in  %
transparent medium), while at long distancies  %
(comparable with \,$\lambda$\,)  the decrease is much more fast  %
(similarly to light in absorbing medium). 

Analogously, in the light of other higher-order DVR   %
it is clear that all scalar components of all higher-order CFs  %
must obey, under BGL,  the same behavior as in (\ref{dvr1l}): 

\begin{eqnarray}
C_k^{0\dots 0}\,\equiv\,  %
[\,\prod_j\, \frac 1{4\pi} \oint_j  \,]\,  C_{k}\,   \Rightarrow\,   %
\, \nonumber \\  \, \Rightarrow\, \,\,  %
[\, \prod_j\, \frac {a^2}{|\rho_j|^2}\,]\,   %
S_k^{0\dots 0} (|\rho_1|/\lambda,\,\dots\,  %
|\rho_k|/\lambda)\,  \,\, ,\, \label{dvrkl} %
\end{eqnarray} 

\  with  \,$\oint_j \,   \dots \,=\, \int d\Omega_j\, \dots$\,  and 
functions \,$S_k^{0\dots 0}$\, scaled by  \,$\lambda$\,   %
independently on  \,$a/\lambda \rightarrow 0$\,. 

\,\,\, 

{\bf 3}.\, Further, let us consider dipole components of CFs  %
and show that under BGL  their dependence on \,$\rho_j$\,s  %
also takes form like (\ref{dvr1l}) and (\ref{dvrkl}). 

From the second  of Eqs.\ref{cn} one has 

\begin{eqnarray}
\partial_t\, C_1^0\,=\,-V\cdot\nabla_R\, C_1^0 \,+   \,  
\label{cn10}  \\  \,+\,  %
(V-v_1)\cdot \frac 1{4\pi} \oint_1  \nabla_{\rho_1}\, C_1\, +  %
\nonumber  \\ +\,   
 \frac 4{\lambda} \int dp_2 \,\,    %
G_m(p_2)\, (v_{2}-V) \cdot   %
 C_{2 } ^{01}( |\rho_2|=a)\,\,\,\, ,\,  \nonumber
\end{eqnarray}

\  where 

\begin{equation} 
C_2^{01} \,=\, \frac 1{4\pi} \oint_1  %
 \frac 1{4\pi} \oint_2 \Omega_2 \, C_2 \,\, \,  \label{c01}
\end{equation}

Integral in the second row here can be transformed  %
with the help of general easy provable identity 

\begin{eqnarray}
  \oint  f(\Omega) \, [\,-\,  (u\cdot \nabla_{\rho})\, C\,]\,   %
  =\, \label{ffo}  \, \\ \,=\, -\,  
  \left(  \frac 2{|\rho|} \,+\, \partial_{|\rho|}\,\right)    %
  \oint (u\cdot\Omega)\,f(\Omega)\, C\,\,+\,  \nonumber \\ \,+\, 
 \frac  1{|\rho|}   \oint  C\,  (u\cdot [\,1 - \Omega\otimes\Omega\,]\,  %
 \cdot\nabla_\Omega)\, f(\Omega)\,\,\,, \,  \nonumber    
\end{eqnarray}

\  with arbitrary function \,$C=C(\rho)=C(|\rho|,\Omega)$\,  %
and \,$\otimes$\, denoting direct (tensor) product of vectors.  %
It yields  

\begin{eqnarray}
\partial_t\, C_1^0\,=\,-V\cdot\nabla_R\, C_1^0 \,- \,  
\label{cn11}  \\  \,-\,  %
 \left(  \frac 2{|\rho_1|} \,+\, \partial_{|\rho_1|}\,\right)  \,  %
(u_1\cdot  C_1^1)\, +  %
\nonumber  \\ +\,   
 \frac 4{\lambda} \int dp_2 \,\,    %
G_m(p_2)\, (v_{2}-V) \cdot   %
 C_{2 } ^{01}( |\rho_2|=a)\,\,\,\, ,\,  \nonumber
\end{eqnarray}

\  where the dipole   \,$C_1$\,'s  component appears, 

\[
C_1^1 \,=\, \frac 1{4\pi} \oint_1  \Omega_1 \, C_1\,\,  %
\]

\  (which is a vector function, naturally).  %

Now, we have to discuss possible   %
dependence of this component on  \,$|\rho_1|$\,  %
at distance \,$|\rho_1|$\, comparable with \,$a$\,.  %
There we can  write  %
\,$C_1^1 = C_1^1(|\rho_1|/a)$\,,  with \,$C_1^1(1)$\,  %
staying finite under BGL, of course.  
Evidently, then second-row expression in Eq.\ref{cn11}   %
stays finite too if and only if 

\[
 [\,  \frac 2{q_1} \,+\, \partial_{q_1}\,]  \,  %
 C_1^1(q_1)\, =\,0\,\,\,  %
\]

\  (otherwise it would tend to infinity \,$\propto \lambda/a$\,). %
This just means that on the whole 

 \begin{eqnarray}
C_1^1\, \Rightarrow\,   %
\frac  { S_1^1 (aq_1/\lambda)}{q_1^2}\,=\,  %
\frac {a^2}{|\rho_1|^2} \, S_1^1(|\rho_1|/\lambda) \,   %
\, \,, \, \label{s11}   
\end{eqnarray} 

\  similarly to (\ref{dvr1l}) (simultaneously,  %
this is confirmation of the BGL asymptotics (\ref{dvr1l})).  %

Hence, Eq.\ref{cn11} turns, - after multiplying it by \,$q_1^2$\,   %
and going to BGL, - into 

\begin{eqnarray}
\partial_t\, S_1^0\,=\,-V\cdot\nabla_R\, S_1^0 \,- \,  
\label{sn10}  
\partial_{|\rho_1|}  \,  %
(u_1\cdot  S_1^1)\, +  \,  \\ +\,   
 \frac 4{\lambda} \int dp_2 \,\,    %
G_m(p_2)\, (v_{2}-V) \cdot   %
 S_{2 } ^{01}( |\rho_2|=0)\,\,\,\, ,\,  \nonumber
\end{eqnarray}

\  where, clearly,\,  \,$S_{2 } ^{01}\,=\, q_1^2\, q_2^2\,  C_{2 } ^{01}$\,.  %

\,\,\, 

{\bf 4}.\, Next, let us discuss evolution equations  %
for  \,$C_1^1$\, and similar dipole or ``multi-dipole''  components  %
of higher-order CFs. 

Applying the identity (\ref{ffo}), with \,$f(\Omega)=\Omega$\,,   %
to some of the space angles, one easy obtains 

\begin{eqnarray}
 \frac 1{4\pi} \oint  \Omega \, [\,-\,  (u\cdot \nabla_{\rho})\, C\,]\,   %
  =\, \label{fo1}  \, \\ \,=\, -\,  
  \frac 13\,u\,\partial_{|\rho|}\, C^0\,-\,  
  [\, \frac  3{|\rho|} \,+\, \partial_{|\rho|}\,]\,    %
  (\, C^2\cdot u\,)\,\,\,,\,  \nonumber 
 \end{eqnarray}
 
 \  where, as above, \,$C^0$\, symbolizes  %
 scalar component of  \,$C=C(|\rho|,\Omega)$\,     %
 (in respect to given space angle), while \,$C^2$\, its  %
 ``quadrupole'' component as defined by 
 
\begin{eqnarray}
  C^2\,=\, \frac 1{4\pi}   \oint   %
  [\, \Omega\otimes\Omega \,-\,\frac 13\,]\,  %
  C\,\,\, \label{quad}     
\end{eqnarray}

\  (thus it is tensor quantity). %

In view of what we already know about behavior of scalar and  %
dipole \,$C_k$\,'s components under BGL,   %
it is obvious that differentiation in  %
the first right-hand term in (\ref{fo1})  %
produces extra factor  %
\,$\propto \lambda/|\rho| \sim  \lambda/a$\, which  %
tends to infinity and therefore must be compensated  %
by proper contribution from the second term.  
The latter, besides, should not produce its own such factor.  %
These requirements mean that the quadrupole component  %
(\ref{quad}) looks like 
\begin{eqnarray}
  C^2\,=\,  \frac {a^2}{|\rho|^2} \, S^2\, +\,  
  \frac {a^3}{|\rho|^3} \, U^2\,\,\,,\,   \label{quad_}     
\end{eqnarray}
\  where \,$S^2=S^2(|\rho|/\lambda)$\, and  %
\,$U^2=U^2 (|\rho|/\lambda)$\,  are scaled  %
by BP's free path (or other value insensitive to BGL),  %
and  \,$S^2$\, is related to \,$S^0$\, by condition 
\begin{eqnarray}
   S^2\, =\,  \frac 23 \,S^0 \,+\,   %
    [\,\frac {u\otimes u}{|u|^2} \,-\,1\,]\,      %
S^{0\prime}\,\,\,, \,   \label{quadc}     
  \end{eqnarray} 
\  which just ensures finiteness of the expression (\ref{fo1})  %
 at  \,$a/\lambda \rightarrow 0$\, and  %
 \,$|\rho|$\, comparable with   \,$a$\,. 

\,\,\, 
  
  Then, notice  that actually the second term  in (\ref{quad_})  %
must have zero value, i.e. \,$U^2=0$\,,   %
since otherwise scalar and dipole components of CFs  %
also would acquire, - through Eqs.\ref{cn}, - 
contributions \,$\propto  a ^3/|\rho_j|^3$\,,  %
which however certainly are forbidden by our  %
previous analysis. 
Besides, factual contribution to \,$C^2$\, from  %
the second term of (\ref{quadc}) equals to zero (merely by definition  %
of this term). 

Due to these  reasons, we find from   (\ref{fo1}) ,  %
(\ref{quad_}) and (\ref{quadc})  that  %
expression (\ref{fo1})  reduces simply  to 

\begin{eqnarray}
 \frac 1{4\pi} \oint  \Omega \, [\,-\,  (u\cdot \nabla_{\rho})\, C\,]\,   %
  =\, \label{fo2}  \, \\ \,=\, -\,  
  u\,\frac  {a^2}{|\rho|^2} \, \partial_{|\rho|}\, S^0\, \,\,   \nonumber
 \end{eqnarray}

Consequently,  equations of evolution of   %
\,$C_k$\,'s  dipole components involve respective \,$C_k$\,'s  %
scalar components only. In particular, we have 

\begin{eqnarray}
\partial_t\, S_1^1\,=\,-V\cdot\nabla_R\, S_1^1 \,- \,  
\label{sn11}  %
u_1\, \partial_{|\rho_1|}  \,  %
  S_1^0\, +  \,  \\ +\,   
 \frac 4{\lambda} \int dp_2 \,\,    %
G_m(p_2)\, (v_{2}-V) \cdot   %
 S_{2 } ^{11}( |\rho_2|=0)\,\,\,\, \,  \nonumber
\end{eqnarray}

Similarly,  evolution of  higher-order  ``multi-scalar-dipole''   %
\,$C_k$\,s components, 

\[
S_k^{\sigma_1\,\dots\,\sigma_k} \,\,\,\,\,\,  %
(\,\sigma_j\,=\,0\,,\,1\,) \,\,\,, \,
\]

involves the same set of functions only, with various   %
``scalar-dipole''  superscript replacements  %
\,$\sigma_j  \Leftrightarrow  1-\sigma_j $\,,  %
plus half of analogous next-order set,  %
\,$S_{k+1}^{\sigma_1\,\dots\,\sigma_k\,1} \,$.  %

\,\,\,  

{\bf 5}.\,  Introducing column 4-vector   %
 \,$\,S_1=\{S_1^0,S_1^1\}$\,, we can unify   %
Eqs.\ref{sn10}  and \ref{sn11}  into 

\begin{eqnarray}
\partial_t\, S_1\,=\,-V\cdot\nabla_R\, S_1 \,- \,  
\label{sn101}  %
U_1\, \partial_{|\rho_1|}  \,  %
  S_1\, +  \,  \\ +\,   
 \frac 4{\lambda} \int dp_2 \,\,    %
G_m(p_2)\, \{0,u_2\} ^\dagger \cdot   %
 S_{2}( |\rho_2|=0)\,\,\,\, ,\,  \nonumber
\end{eqnarray}

\  with 4$\times$4\,-matrix 

\[
U_1\,=\, \left\{\,  %
\begin{array} {cc}  %
 0 &  u_1^\dagger   \\   %
u_1  &    0 
\end{array} 
\right\}\,\,\, , \,\, 
\] 

\  symbol   \,$\dagger$\, denoting vector or matrix transposition,  %
and row 4-vector \,$\{0,u_2\} ^\dagger$\,   %
associated with ``outer'' atom. 

Quite similarly, all \,$2^k$\, scalar and dipole components   %
\,$S_k^{\sigma_1\dots \sigma_k}$\, can be unified into single  %
\,$4\times \dots\times 4$\,-tensor object \,$S_k$\,,  
and then all evolution equations for these components  %
replaced by more compact hierarchy of equations for  %
tensors \,$S_k$\, which trivially generalize Eq.\ref{sn101}. 


\,\,\, 

{\bf 6}.\,  Thus, it seems that  Eq.\ref{sn101}   %
altogether with its just mentioned higher-order  %
analogues (plus first of Eqs.\ref{cn} for \,$S_0\equiv C_0$\,)   %
form a closed (although infinite) system of equations .  %

In fact, however, situation is not so comfortable.    %
The matter is that at the same time the collision boundary conditions  %
(CBC),  (\ref{crc}) or    (\ref{rc}),   %
   in general involve also \,$S_1^2$\, and various   %
other quadrupole and multi-pole components of BP-atoms   %
correlations (corresponding to \,$\sigma_j =2,3,\,\dots \,$). 
For example,  at \,$k=1$\, our CBC (\ref{rc}), after its multiplying  %
by \,$f(\Omega_1)=1$\, or \,$f(\Omega_1)=\Omega_1$\,  %
and space-angle averaging produces  

\begin{eqnarray}
(\nabla_{p_1} -\nabla_P) \cdot   %
S_1^1(|\rho_1| =0,\,P,p_1) \,=\,0\,\,\,, \,\,\, \,\,\,\,\, \label{arc11}  \\ 
\frac 13\, (\nabla_{p_1} -\nabla_P) \,   %
[\,S_1^0(|\rho_1| =0,\,P,p_1)\,+\,S_0(P)\,]\, +\,\,  \,\, \nonumber  %
\\ \,+\,  %
(\nabla_{p_1} -\nabla_P) \cdot  %
S_1^2(|\rho_1| =0,\,P,p_1)\,=\,0\,\,\,, \, \,\,\,\, \,\,\,\,   \label{arc102} 
\end{eqnarray} 

\  where \,$S_0=C_0$\,, and we took into account  %
our above analysis. 
At that,  there are no evident formal reasons   %
 to exclude second term of expression (\ref{quadc}),  %
e.g. assuming \,$S_1^{0\prime}=0$\,,  %
and thus reduce  \,$S_1^2$\, to \,$S_1^0$\,.  
 
Hence, in contrast to evolution equations themselves,  %
the CBC they require are not closed in respect to   %
scalar and dipole CF's components, even under BGL. 

\,\,\, 

 Nevertheless,  just presented  consideration    %
may be base for a meaningful approximation   %
in Eq.\ref{arc102} and similar  CBC   %
and thus for approximate solution of  Eqs.\ref{cn}. 

All that is interesting subject  for separate discussion.  %
 Here, we at least illustrated by one more method that  %
BGL does not ``lighten'' the problems of gas kinetics.   %
They remain, again resembling ``Cheshire Cat's smile''.  %


\subsection{Pseudo-Liouville  %
representation of hard-ball dynamics}

{\bf 1}.\, In this representation of the conventional   %
theory (see e.g, \cite{pg})  %
the CBC (\ref{ccrs}) are directly inserted into Liouville operator  %
in the form of singular ``interaction'' term.  
In application to our particular system ``BP+atoms''   %
this formal trick means that 
 
\begin{eqnarray}
\lo_x\, \Rightarrow\, -u\cdot\nabla_\rho \,+\, \,\, \nonumber   %
 \\ \,+\,    %
a^2 \oint \delta(\rho -a\Omega)\,\,  %
(\Omega\cdot u)\, \widehat{S}(\Omega,P,p)\,\,\,, \,\,\,\,  \label{hlo} 
\end{eqnarray}

\ where, as above, \,$u=v-V$\,, and  %
\,$\widehat{S}(\Omega,P,p)\,$ is operator defined by 

\begin{eqnarray}
\widehat{S}(\Omega,P,p)\,F(P,p)\,=\,\, \,\,\,\,  \,\,\,  \label{os} \\ \,=\, 
\theta(-\Omega\cdot u)\,F(P,p)\,+\,   %
\theta(\Omega\cdot u)\,F(P^*,p^*)\,\,\, \,\,\,\,  \,\,\nonumber
\end{eqnarray}

\  Here \,$\theta(\cdot)$\, is Heaviside step function,  %
and \,$P^*$\, and   \,$p^*$\, are pre-collision momenta  %
corresponding to post-collision  \,$P$\, and  %
 \,$p$\,  in accordance with  relations (\ref{un})-(\ref{ut}). 
At that, the relative distance \,$\rho$\, formally gets rights to take  %
values from physically forbidden regions,  while 
\[
 g(x)\,\Rightarrow\,\theta(|\rho|-a)\, G_m(p)\,\, 
\]

Inserting expression (\ref{hlo})  into Eqs.\ref{fo} and \ref{fop},   %
we obtain related generating-functional evolution operators  %
\,$\fo$\, and \,$\fo^\prime$\,. 
They are equivalent to full hard-ball BBGKY hierarchy  considered  %
in terms of DFs \,$F_k$\, and CFs \,$C_k$\,, respectively. 

\,\,\, 

{\bf 2}.\,  One of formally significant differences of this  %
approach to hard-ball  %
limit from above discussed case of smooth interaction  %
is that now  \,$\fo^\prime$\, contains non-zero  %
 \,\,$\psi(x)$\,-independent part,   %
in contrast to (\ref{fop}). Concretely, 

\begin{eqnarray} 
\fo^\prime\{V,\psi =0,\nabla_P,\delta/\delta\psi =0\}\,=\,  \,\,\, \nonumber  %
\\  \,=\,  %
\int_x\, n\,  \lo_x\, g(x)\,\,=\, \bo\,\,\neq 0\,\,\,, \,\,\,  \label{hbo} 
\end{eqnarray}

\  where \,$\bo$\, is already mentioned  Boltzmann-Lorentz   %
operator (BLO). Now,  according to  (\ref{hlo}) and (\ref{os}), %
its action is described by 

\begin{eqnarray} 
\bo\,\Rightarrow\,n\, %
a^2\int_p \oint   %
(\Omega\cdot u)\,   %
\,\times \,\,\,  \nonumber \\ \,\times\,  %
[\, \widehat{S}(\Omega,P,p) \,   %
-\,1\,]\, G_m(p)\,\,\,, \,\,\,\,  \label{hbo_} \\
\bo\,F(P)\,=\, n\,a^2\int_p \oint  (\Omega\cdot u)\,  %
\theta(\Omega\cdot u)\,    %
\,\times \,\,\,  \nonumber \\ \,\times\,  %
[\,G_m(p^*)\,F(P^*)\,-\,G_m(p)\,F(P)\,]\,\,\, \,\,  \nonumber
\end{eqnarray}

Formally, this part of \,$\fo^\prime$\, results from the lower-order   %
terms of the hard-ball collision boundary conditions (CBC) ,  %
Eq.\ref{crc} (Eq.13 in \cite{hs}), as clearly shows  %
the integrand expression   in Eq.17 in \cite{hs}. 


\subsection{Pseudo-kinetic  %
representation of hard-ball dynamics}

{\bf 1}.\, Physically, the BLO part of \,$\fo^\prime$\,  %
can be interpreted as representative of equilibrium thermostat  %
consisting of particles (atoms) uncorrelated with %
our ``Brownian particle'' (BP) under observation. 
Separating this part evidently, let us  write

\begin{eqnarray}
\fo^\prime \,=\,\bo\,+\,\fo^+ \, +\, \fo\,\,\,, \,\,\,  
 \label{hfop} \\    \nonumber   %
\fo^+\,\equiv\,  %
\int_x \psi(x)\, \lo_x\,g(x)\,  \,\, \,\,   
\end{eqnarray} 

Here, we separated also \,$\fo^\prime$\,'s component   %
\,$\fo^+$\,  which rises \,$\psi(x)$\,-dependence  %
of an operand (like ``creation operator'',  %
in the sense of \cite{p0806,last}),   %
while two components of \,$\fo$\,, - namely,      
\begin{eqnarray}
\fo\,=\,\fo_0\,+\, \fo^- \,\,\, ,  %
 \,\,\,\,  \,\,\, \,\,\,  \,\,\,   \,\,\,  \label{hfo}  \\  \nonumber  %
\fo_0\,\equiv \int_x \psi(x)\, \lo_x\, \frac \delta{\delta \psi(x)} \,\,, \,\,\,\,   %
\fo^-\,\equiv \, n \int_x \lo_x\, \frac \delta{\delta \psi(x)} \,\,\,, \,
\end{eqnarray} 

- respectively conserve this dependence on \,$\psi(x)$\,    %
and lower it (like ``annihilation operator''). 
 
Then, evolution equation for  CF's generating   %
functional (GF)  can be rewritten as 
 
\begin{eqnarray}
\partial_t\, \mathcal{C}\,=\,[\,-V\cdot\nabla_R\,+  %
\bo +\fo^+ +\fo\,]\, \, \mathcal{C} \,\,\,, \, \,   \label{hcfe_}   %
\end{eqnarray}  
 
with the same initial condition (\ref{fcic}), as before,  %
 and same relation to BP's probability distribution function,  

\[
 F_0(t,R,P)\,=\,  \mathcal{C}\{\psi =0\}\,\, 
\]

{\bf 2}.\,  If  random walk of hard-ball  BP    %
obeyed Boltzmann's molecular chaos,  %
then the generating-functional part of evolution operator  %
in Eq.\ref{hcfe_}, - i.e. two last terms of (\ref{hfop}), -  %
would have no effect onto \,$F_0(t,R,P)$\, at least  in  BGl. 
In order to highlight the question, - is it really so or not, -   %
it is convenient again to represent Eq.\ref{hcfe_},  -  %
along with equivalent Eq.\ref{gfe_} for \,$\mathcal{F}$\,,  -  %
via ``pseudo-kinetic'' operators. 

Evidently,  such transform by itself does not differ from what we  %
made in case of smooth interaction potential. Therefore,  %
repeating derivation of  Eqs.\ref{kc}-\ref{kop1}, we again can  write  

\begin{eqnarray}
\partial_t\, \mathcal{C}\,=\,-V\cdot\nabla_R\,\mathcal{C} \,+      %
\,\ko^\prime  (t)\, \mathcal{C} \,\,\,, \,  \label{hkc}
\end{eqnarray} 

\ with  ``pseudo-kinetic'' operator (PKO) 

\begin{eqnarray}  
\ko^\prime (t)\,=  \int_x\, [\,n+\psi(x)]\, \lo_x \,  \,\times\,  %
\,\,\,\,\,\,\,\,\, \label{hkop}   \\  \,\times \, \, 
\exp{[(\lo_x + \fo_R^\prime )\,t\,]}\, g(x)\, \exp{[\,-  %
 \fo_R^\prime \,t\,]}\, \, \, \nonumber 
\end{eqnarray} 

\  or, equivalently, 

\begin{eqnarray}  
\ko^\prime (t)\,= 
 \int_x [\,n + \psi(x)]\, \lo_x\,   g(x)\, \,+\,\,\,\,\,\,\,  %
\,\,\,\, \label{hkop1}  \\ \,+ 
\int_x [\,n+\psi(x)]\, \lo_x   \int_0^t d\tau\,  %
\,\times\,\,\, \nonumber \\ \,\times \,\,  %
\exp{[(\lo_x + \fo_R^\prime )\,\tau\,]}\, \lo_x\,  g(x)\,  %
\exp{[\,-  \fo_R^\prime \,\tau\,]}\, \, \, \, \,   \nonumber 
\end{eqnarray} 

\   Here, as before, \,\,$ \fo_R^\prime = -V\cdot\nabla_R +  %
 \fo^\prime \,$.  %
Notice that in the last transition from Eq.\ref{hkop}  %
to Eq.\ref{hkop1}, in contrast to transition  %
from  Eq.\ref{kop}  to Eq.\ref{kop1}, we have not removed term  %
\,\,$n\int_x \lo_x\,g(x)$\,, since now, - according to   %
Eq.\ref{hbo}, - it does not turn to zero. 

The latter formula can be rewritten also as 
\begin{eqnarray}  
\ko^\prime (t)\,=\, \bo\,+\,\fo^+\,   \,+\,\,\,\,\,\,\,  %
\,\,\,\, \label{hkop_}  \\ \,+ 
\int_x [\,n+\psi(x)]\, \lo_x   \int_0^t d\tau\,  %
\,\times\,\,\, \nonumber \\ \,\times \,\,  %
\exp{[(\lo_x + \fo_R^\prime )\,\tau\,]}\, \lo_x\,  g(x)\,  %
\exp{[\,-  \fo_R^\prime \,\tau\,]}\, \, \,, \, \,   \nonumber 
\end{eqnarray} 
- to be combined with Eqs.\ref{hfop} and \ref{hfo}.   %

Then, solution of Eq.\ref{hkc}, in respect to   %
BP's marginal distribution, can be represented by 
\begin{eqnarray}  
F_0(t,R,P)\,=\,  \,\,\,\,\,\,\,\,\,\,\,   \label{hke} \\  =\,  %
\left\langle\, \overleftarrow{\exp}\,  %
\int_0^t [-V\cdot\nabla_R + \ko^\prime(t^\prime)\, ] \,   %
dt^\prime\, \right\rangle \,\times\, \nonumber \\   %
\,\times \,\, F_0(0,R,P)\,\,\, ,\,\, \nonumber  
\end{eqnarray} 

\ where angle brackets mean, as before, statistical average  %
over ensemble of dynamical trajectories of the system  %
(BP + gas) weighted by equilibrium distribution of  %
initial gas microstates. In our designations, 
 \[
 \langle\, \dots\,\rangle \,=\, [\,\dots\,]_{\psi =0} \,\,\, 
\]
In the language of quantum field theory \cite{p0806,last}),  %
this is ``vacuum average'' \,$\langle 0|\dots |0\rangle$\,, %
or amplitude of vacuum-vacuum transition, where role  %
of ``vacuum'' is played by gas equilibrium. 

\,\,\, 

\subsection{Failure of Boltzmannian kinetics   %
under hard-ball Boltzmann-Grad limit}

{\bf 1}.\, The last two  right-hand terms in Eq.\ref{hkop_}  %
represent contribution to BP's kinetics from gas correlations  %
with BP's random walk. In general, undoubtedly, they  %
give substantial addition to the Boltzmann's term \,$\bo$\,,  %
both in the sense of ensemble average  and fluctuations  %
of BP's relaxation rates (``friction'', ``diffusivity'', ``mobility'', etc.).  
Therefore, let us go to BGL. 

This purpose again can be achieved with the help of scale   %
transformation like (\ref{st}), namely, 

\begin{eqnarray}  
n\,\Rightarrow\, \frac n{\xi^2}\,\,\,, \,\,\,\,\, %
a\,\Rightarrow\,  \xi\,a\,  \,\,, \,\,\, \label{hst} %
\end{eqnarray} 

\  where\, \,$\xi\,\rightarrow\,0$\,, - with  %
\,$a \Rightarrow  \xi \,a\,$ applied in the singular ``collision''  %
term of pseudo-Liouville operator (\ref{hlo}), -  
supplemented by the same changes of variables  %
 \,$x=\{\rho,p\}$\,, \,$\psi(x)$\, and   \,$\tau$\,  %
as in (\ref{st_}),   %
and same transformation of the variational derivative   %
as in (\ref{st__}). 

At that, at\,  \,$\xi\rightarrow 0$\,\, we have   %
\,$\lo_x\Rightarrow \xi^{-1}\lo_x$\,, and operators \,$\fo$\,   %
(\ref{hfo}) and  \,$\fo^\prime$\, (\ref{hfop})  transform  %
asymptotically exactly as in Eq.\ref{lfo}.  
Consequently, we come to    

\begin{eqnarray}  
\ko^\prime (t)\,\,\Rightarrow\,  \,    \ko_\infty\,=\, 
\bo\,+\,\fo^+\, +\, \, \,\,\,\,\, \, \label{hlko}  \\ \,+ 
\int_x [\,n+\psi(x)]\, \,\lo_x   \int_0^\infty  d\tau\,  %
\,    \times\,\,\, \nonumber \\ \,\times \,\,  %
\exp{[(\lo_x + \fo )\,\tau\,]}\, \lo_x\,  g(x)\,  %
\exp{[\,-  \fo \,\tau\,]}\, \, \, \, \,   \nonumber 
\end{eqnarray} 

\,\,\,

Hence, the exact pseudo-kinetic operator does not  %
reduce to Boltzmann-Lorentz operator (BLO)  %
even under Boltzmann-Grad limit (BGL). 

 Let us show that  same statement is true also  %
in respect to result of statistical averaging in the  %
exact Eq.\ref{lke} following from Eq.\ref{hke} under BGL,  %
that in respect to actual statistics of random walk of our   %
Brownian particle (BP). 

\,\,\,

{\bf 2}.\, Basing on logical necessity of correspondence between  %
cases of hard-ball and smooth interactions, we expect  %
that now also on average 
\[
 \langle\,\ko_\infty\,\rangle\, =\, \bo\,\,\, ,\,\, 
\]
that is average value of the integral term in Eq.\ref{hkop_}  %
equals to zero. It is really so. 

Indeed, since action of \,$\fo$\,  %
onto any \,$\psi(x)$\,-independent object produces zero,  
we have  
\begin{eqnarray}  
 \langle\,\ko_\infty\,\rangle\, =\, \bo\,+\, 
%
n  \int_0^\infty  d\tau \int_x \lo_x \,  %
 e^{\lo_x \,\tau}\, \lo_x\,  g(x)\,  \, \, \,   \nonumber 
\end{eqnarray} 
The integral here by its exterior looks like in     %
the ``smooth case'', but now its contents  %
represents not a single BP-atom collision  %
but two consecutive  BP's collisions with one  %
and the same atom. 
Since such event is kinematically  impossible,  %
we can state that this integral is equal to zero,  %
and therefore \,$\langle \ko_\infty\rangle =\bo$\,. 

\,\,\, 

{\bf 3}.\, Next, consider variance of the limit   %
pseudo-kinetic operator \,$\ko_\infty\,$. 

Evidently, first, we can write 
\begin{eqnarray}  
 \langle\,(\ko_\infty  -  \langle\ko_\infty\rangle )^2 \,\rangle\,   %
\,=\,\,\,\,\, \, \,\,\,\, \label{lcov}  \\  \nonumber  =\,  
\left\langle n\int_x \lo_x   \int_0^\infty \!\!  d\tau\,  %
e^{(\lo_x + \fo )\,\tau}\, \lo_x\,  g(x)\,  %
e^{ - \fo\tau}\,\fo^+\, \right\rangle \, 
\end{eqnarray} 

Second, in expansion of the exponentials there  %
over ``annihilation operator''  \,$\fo^-$\,,  %
 according to expansion (\ref{hfo}),  %
 only first-order term  survives after averaging, so that 
\begin{eqnarray}  
 \langle\,(\ko_\infty  -  \langle\ko_\infty\rangle )^2 \,\rangle\, =\,    %
   \label{lcov1}  %
n\int_x \lo_x   \int_0^\infty  \!\! d\tau  %
\int_0^\tau  \!\! d\eta\,  
\, \times\, \,\,  \\  \,\times \,\,  \nonumber
\langle\,   [\, e^{\lo_x \,(\tau -\eta)}\,\fo^- \,  %
e^{(\lo_x + \fo_0)\,\eta}\, \lo_x\,  g(x)\,   %
e^{ - \fo_0\tau}\,\,-\, 
 \, \nonumber \\ \,- \,\,  
e^{\lo_x\,\tau}\, \lo_x\,  g(x)\,  %
\fo^-\, e^{ - \fo_0\,\eta}\,] \, \,\fo^+\, \rangle \, \,\,\, \nonumber
\end{eqnarray} 

Third, substituting there  \,$\fo^-$\,,  \,$\fo_0$\, and  %
\,$\fo^+$\, from   (\ref{hfo})   and  (\ref{hfop}),   %
we come to visually same expression as in  (\ref{kov}). 
For convenience, we write out it repeatedly: 

\begin{eqnarray}  
\langle\,(\ko_\infty  -  \langle\ko_\infty\rangle )^2 \,\rangle\, =\,   %
n^2 \int_x \int_y  %
\int_0^\infty  \!\! d\tau  \int_0^\infty \!\!  d\tau^\prime\, \,  %
\times \,\,\,\, \,\,\,\,\,  \label{hkov} \\ \times\,\,    \nonumber 
\lo_x\,[\, e^{\lo_x\,\tau}\,\lo_y\,  %
e^{(\lo_x +\lo_y)\, \tau^\prime\,} \, \lo_x\,  %
e^{-\lo_y\,(\tau +\tau^\prime\,)}\,   %
\, -\, \,\, \\ \,-\,   
e^{\lo_x\,(\tau +\tau^\prime\,)}\, \lo_x\,  %
\lo_y\, e^{-\lo_y\,\tau^\prime\,}\,]\,  \lo_y\,g(y)\,g(x)\,\,\,  \,\,  \nonumber
\end{eqnarray} 

Its difference from  (\ref{kov}) is determined by that of    %
singular pseudo-Liouville \,$\lo_x$\, (\ref{hlo}) from ``smooth''  %
Liouville operator (\ref{lo}). 

\,\,\,

{\bf 4}.\,  Figuratively speaking, each of most left or most right-hand   %
of operators \,$\lo_x$\, and \,$\lo_y$\, in Eq.\ref{kov} is  %
responsible for end or beginning of same collision, respectively, i.e.  %
``half of collision''. 
What is for Eq.\ref{hkov}, in opposite, one can say that there each   %
of these operators represents complete separate collision, since  %
the singular \,$\delta$\,-function part of pseudo-Liouville  %
operator (\ref{hlo}) by itself makes it. 

Under such treatment,  second and third rows in Eq.\ref{hkov}   %
describe two variants of four BP's collisions with two atoms.  

At that, clearly, events corresponding to the third row  %
in fact can not realize, by the same kinematic reasons  %
by which the above considered integral in  %
\,$\langle\ko_\infty\rangle $\, turns to zero.  %
Namely, because two directly consecutive collisions  %
between (mutually repulsing) particles are impossible.  %

Therefore, we can remove the third row and rewrite Eq.\ref{hkov}  %
simply as
 
\begin{eqnarray}  
\langle\,(\ko_\infty  -  \langle\ko_\infty\rangle )^2 \,\rangle\, =\,   %
n^2 \int_x \int_y  %
\int_0^\infty  \!\! d\tau  \int_0^\infty \!\!  d\tau^\prime\, \,  %
\times \,\,\,\, \,\,\,\,\,  \label{hkov_} \\ \times\,\,    \nonumber 
\lo_x\, e^{\lo_x\,\tau}\,\lo_y\,  %
e^{(\lo_x +\lo_y)\, \tau^\prime\,} \, \lo_x\,  %
e^{-\lo_y\,(\tau +\tau^\prime\,)} \,  \lo_y\,g(y)\,g(x)\, \,\,\,   %
\end{eqnarray} 

Formally,  this expression   %
corresponds to four alternate BP's collisions with two atoms  %
(described by \,$x=\{\rho_x,p_x\}$\, and   %
\,$y=\{\rho_y,p_y\}$\, ). %
The alternation is important there, because just it   %
makes such events kinematically allowed and possible. 

Moreover, alternation  ensures kinematic  %
and dynamic possibility of arbitrary large  %
number of collisions between BP and two atoms,  %
and such complicated events also are covered    %
by Eqs.\ref{hkop}-\ref{hkop_},   %
due to presence of the singular \,$\lo_x$\,'s  and 
\,$\lo_y$\,'s parts, - in the role of ``kinetic operators'', -   %
in the exponentials there. 

Hence, we can state that the operator variance  %
(\ref{hkov})-(\ref{hkov_}) is not zero.  
This means that Boltzmannian kinetics fails,   %
and the exact kinetic (``pseudo-kinetic'') operator  %
stays different from the Boltzmann's one and random  %
even in case of hard-ball interaction  even under BGL!   

This follows also from the ``correspondence principle ''  %
and above similar statement  for arbitrary (in particular,  %
arbitrarily sharp) smooth interaction. 

\,\,\, 

{\bf 5}.\,  Physically, however, too literal treatment of  %
Eqs.\ref{hkop}-\ref{hkop_}  in terms of multiple collisions  %
is rather incorrect. 
 
We should not forget that integrand  there   %
is nothing but  (second-order) statistical moment of random  %
(operator-valued) quantity.  

Therefore seemingly repeated collisions by essence  %
may belong to different ``stories'', each without repetitions,   %
with physically different particles. 
At that, repetitions are merely synonyms of kinematic and  %
dynamical intersections  between possible   %
variants of system's evolution. 
This is seen from  %
the seed form (\ref{hkop}) of the limit pseudo-kinetic  %
operator (\ref{hlko}), which displays just interference of current collision   %
in microstate evolution of the rest of gas,  %
or, reciprocally,  interference of the latter in realization of the former.    %

Though, on the other hand, repeated (alternated) collisions and multiply   %
repeated ones in hard-ball systems is their  immanent specificity,   %
supported by zero duration of individual collision.     %
Therefore, to some extent, they play role of ``smooth''    %
time-stretched collisions. Under such interpretation, for example,   %
the expression inside angle brackets in Eq.\ref{lcov}   %
says about influence of the system's history, i.e. past collisions,    %
on ``probability'' of  present realization of (alternated)   %
repeated collision. 


\section{Conclusion}

{\bf 1}.\, Main goal of this manuscript  was,   %
firstly,  new demonstration,    %
in several original ways, of that the Boltzmann-Grad limit    %
(BGL) does not eliminate effects of  inter-particle statistical    %
correlations and therefore does not lead to the   %
Boltzmann's kinetics.   %
Secondly, this statement equally relates to cases of smooth potential  %
interactions between particles and the hard-ball interaction. 

The correlations do their work though surviving at  %
zero-measure phase-space subsets only,  like  %
``angels on needle tip''. 

\,\,\, 

{\bf{2}.\,  In this sense, BGL does not exist. In fact, it results in  %
a non-trivial non-Boltzmannian kinetics which depends on    %
parameter \,$\pi a^2n =\lambda^{-1}$\, as the whole only, -   %
with \,$a$\,, \,$n$\, and \,$\lambda$\,   %
denoting interaction radius, mean number density   %
of particles and characteristic free-flight length, -   %
but not on  \,$a$\, or  \,$n$\, separately   %
(in presence of different sorts of particles, another essential   %
parameters may be their mass, radius and density ratios). 

At that, statistics of random walk of a probe  %
``Brownian'' particle (BP)  is exactly governed by  %
random ``pseudo-kinetic'' operator (PKO) which takes place of %
Boltzmann or Boltzmann-Lorentz kinetic operator  %
and coincides with it (under BGL) on average only. 

\,\,\, 

{\bf 3}.\,  Randomness (fluctuations) of the PKO   %
reflects hugeness of number of system's initial microstate  %
parameters (variables) determining BP's walk,  %
as compared with number of parameters characterizing this walk. 

Importance of giant difference between these numbers  %
already was pointed out in   %
\cite{p1,tmf} and other our works.  It implies  %
impossibility of time averaging of BP's relaxation rate,  %
i.e. relative frequency and efficiency of its collisions.   %
All the more, because the difference even grows with   %
observation time (and, moreover, turns to infinity  %
under BGL).  In other words, it implies non-ergodicity  %
of kinetic properties of ``molecular Brownian motion'' \cite{tmf}. 

As the consequence, diffusivity and mobility of BP   %
possess no a priori certain value, instead changing  %
unpredictably from one experiment, - that is realization   %
of BP's walk, -  to another. 
Equivalently, we can say that diffusivity (mobility) undergoes   %
scaleless low-frequency fluctuations like 1/f-noise  %
(see references above). Randomness of the PKO  %
just produces such kind of fluctuations

\,\,\, 

{\bf 4}.\, So crucial disagreement between our conclusions and that  %
of the pure mathematical analysis of hard-ball gas under BGL   %
in \cite{cpg,pg} (and earlier in \cite{lan2})  %
is not surprising:\, as we underlined in the body of this manuscript,  %
our consideration concerns events involving finite number of particles   %
but arising in virtually active background of infinitely many   %
other particles, while mathematicians consider events with   %
literally finite particles' numbers. 

To some extent, this resembles quantum mechanics   %
with and without physical  vacuum   %
(and such analogy is quite meaningful, since   %
statistical-mechanical problems  under our attention indeed can   %
be reformulated in terms of quantum field theory  %
\cite{p0806,p1302,last}).  

Besides, mathematicians filter   %
events by estimates and reasonings based on a priory  %
Lebesgue measures, while our approaches reveal   %
a posteriori significant events (confirming sentence   %
``real is unprobable'').  %

Of course, our theory needs in more formal rigor.  %
At the same time, in our opinion, mathematical theory  needs    %
in principal improvements, - in order to become closer to physics, -   %
at that taking into account the Krylov's criticism \cite{kr}   %
and getting rid of ancient  prejudices   %
of  Botzmannian kinetics. 
 
\,\,\,  

{\bf 5}.\,  In the framework of our theory, now new question   %
appears - about inter-relations   %
between earlier suggested approaches to 1/f-noise in  %
diffusivities (mobilities), as well as other relaxation rates,  %
and presently suggested method of the exact  %
random pseudo-kinetic operator. This is one more   %
interesting task for future.





\begin{thebibliography}{41}
\bibitem{fn1}
\,More explanations of ideas of our approaches to investigation  %
of molecular random walks in many-particle systems (fluids),  %
some results of these approaches and their discussions  %
can be found in works \cite{bk3}-\cite{last}  %
(see also references therein). 

\,Their application to specific hard-ball systems  %
were concerned in \cite{hs,hs1,ig}. 

\,Their applications to systems (solids) considered in terms of  %
collective (e.g. phonon) variables and, besides, some %
related abstract formulations and results %
are expounded in \cite{i3}-\cite{p1302} (see also references therein). 

\,\,\,
\bibitem{fn2}
\, For general case of BP in non-ideal gas (or liquid) or %
fluids in themselves,  the cumulant distribution functions and %
related DVR were considered in \cite{p0802,p0803,tmf,p1105,p1203}.

\,\,\,
\bibitem{fn3}
Such ``quantum field-theoretical'' terminology in fact has strong   %
mathematical justification, as it was shown in \cite{p0806,last}.
\end{thebibliography}
\end{document}